\documentclass[twocolumn, dvipsnames]{aastex62}
\submitjournal{ApJ}

\usepackage{amsmath,natbib, float, color, soul, xspace}
\usepackage[normalem]{ulem}
\usepackage{listings}

\newcommand*\Msolarh[0]{h^{-1} \, \mathrm{M_{\odot}}}

\newcommand*\Mpch[0]{h^{-1}\,\mathrm{Mpc}}

\newcommand*\sig{\textsc{$\sigma_8$}\xspace}

\newcommand*\om{\textsc{$\Omega_m$}\xspace}
\newcommand*\Ho{\textsc{$H_0$}\xspace}

\newcommand*\planck{\textit{Planck}\xspace}

\newcommand*\lcdm{$\Lambda$CDM\xspace}

\newcommand*\abacus{\texttt{AbacusCosmos}\xspace}

\newcommand*\new[1]{{#1}}

\newcommand{\code}[1]{\texttt{#1}}

\shorttitle{}
\shortauthors{Ntampaka, et al.}

\begin{document}

\correspondingauthor{Michelle Ntampaka}
\email{michelle.ntampaka@cfa.harvard.edu}

\author{Michelle Ntampaka}
\affiliation{Harvard Data Science Initiative, Harvard University, Cambridge, MA 02138, USA}
 \affiliation{Center for Astrophysics $|$ Harvard \& Smithsonian, Cambridge, MA 02138, USA}

\author{Daniel J. Eisenstein}
\affiliation{Center for Astrophysics $|$ Harvard \& Smithsonian, Cambridge, MA 02138, USA}

\author{Sihan Yuan}
\affiliation{Center for Astrophysics $|$ Harvard \& Smithsonian, Cambridge, MA 02138, USA}

\author{Lehman H. Garrison}
\affiliation{Center for Astrophysics $|$ Harvard \& Smithsonian, Cambridge, MA 02138, USA}
\affiliation{Center for Computational Astrophysics, Flatiron Institute, New York, NY 10010, USA}

\title{A Hybrid Deep Learning Approach to Cosmological Constraints From Galaxy Redshift Surveys}

\begin{abstract}
We present a deep machine learning (ML)-based technique for accurately determining \sig and \om  from mock 3D galaxy surveys.  The mock surveys are built from the \abacus suite of $N$-body simulations, which comprises 40 cosmological volume simulations spanning a range of cosmological models, and we account for uncertainties in galaxy formation scenarios through the use of generalized halo occupation distributions (HODs).  We explore a trio of ML models:  a 3D  convolutional neural network (CNN), a power-spectrum-based fully connected network, and a hybrid  approach that merges the two to combine physically motivated summary statistics with flexible CNNs. We describe best practices for training a deep model on a suite of matched-phase simulations and we test our model on a completely independent sample that uses previously unseen initial conditions, cosmological parameters, and HOD parameters. Despite the fact that the mock observations are quite small ($\sim0.07h^{-3}\,\mathrm{Gpc}^3$) and the training data span a large parameter space (6 cosmological and 6 HOD parameters), the CNN and hybrid CNN can constrain \sig and \om to $\sim3\%$ and $\sim4\%$, respectively. \\ \\
\end{abstract}


\section{Introduction}

\label{sec:intro}

In the \lcdm cosmological model, tiny density fluctuations in the early Universe evolved into today's cosmic web of overdense dark matter halos, filaments, and sheets.  Imprinted on this large-scale structure is information about the underlying cosmological model, provided one knows how and where to look.  Measurements that describe the large scale distribution of matter in the Universe carry information about the cosmological model that drove its formation. These measurements include descriptions of the spatial distribution and clustering of galaxies \citep[e.g.,][]{1990ApJS...72..433H, 1996ApJ...470..172S, 2001MNRAS.327.1297P, 2004ApJ...606..702T},  the abundance of massive galaxy clusters \citep[e.g.,][]{2009ApJ...692.1060V, 2015MNRAS.446.2205M,2016ApJ...832...95D},  the weak gravitational lensing of galaxies by intervening large-scale structure \citep[e.g.,][]{2000MNRAS.318..625B, 2000astro.ph..3338K,  2000Natur.405..143W, 2000A&A...358...30V, 2017arXiv170801530D, 2018arXiv181206076H, 2019PASJ...71...43H}, and the length scale of baryon acoustic oscillations \citep[e.g.,][]{2005ApJ...633..560E, 2005MNRAS.362..505C, 2017MNRAS.470.2617A}.  \new{A hallmark difference between these and probes of the earlier Universe is non-Gaussianity:  though the early Universe is well-described by a Gaussian random field \citep[e.g.,][]{2014A&A...571A..16P, 2014A&A...571A..24P}, gravitational collapse drives the formation of non-Gaussian correlations in the late-time matter distribution.}  See \cite{2013PhR...530...87W} for a review of these and other observational cosmological probes. 

Galaxies live in dark matter halos and are tracers, albeit biased ones, of  large-scale structure.   Large spectroscopic surveys such as the Sloan Digital Sky Survey \citep[SDSS,][]{2000AJ....120.1579Y} have produced maps of the 3D distribution of galaxies in the Universe, and upcoming spectroscopic surveys such as the 
Dark Energy Spectroscopic Instrument \citep[DESI,][]{2016arXiv161100036D},  Subaru Prime Focus Spectrograph \citep[PFS,][]{2014PASJ...66R...1T}, 4-metre Multi-Object Spectroscopic Telescope \citep[4MOST,][]{2014SPIE.9147E..0MD}, and \textit{Euclid} \citep{2013LRR....16....6A} will produce exquisitely detailed maps of the sky.  The galaxy power spectrum provides one handle on summarizing and interpreting these 3D galaxy maps and can be used to put constraints on the parameters that describe a \lcdm cosmology \citep[e.g.,][]{2004ApJ...606..702T}, but care must be taken when disentangling the effects of cosmology and galaxy bias \citep[e.g.,][]{2013MNRAS.430..725V, 2013MNRAS.430..747M, 2013MNRAS.430..767C}.



Though it is an abundantly useful compression of the  information contained in the distribution of galaxies, the power spectrum is not a \textit{complete} accounting of this information because the late-time galaxy distribution is not a Gaussian random field.  The deviations from Gaussian correlations are enormous at small length scales ($\lesssim$ a few Mpc), where dark matter halos have collapsed and virialized, but remain substantial at intermediate scales due to the cosmic web of filaments, walls, and voids. Additional statistics such as the squeezed 3-point correlation function \citep{2018MNRAS.478.2019Y}, redshift space power spectrum \citep{2019arXiv190708515K}, counts-in-cylinders \citep{2019arXiv190309656W}, and the minimum spanning tree \citep{2019arXiv190700989N} have been shown to be rich in complementary cosmological information\new{ by capturing non-Gaussian details of the galaxy distribution that are not described by the power spectrum alone.}  

These higher-order statistical descriptions of how galaxies populate 3D space typically need to be calibrated on cosmological simulations.   Cosmological hydrodynamical simulations that trace the formation of galaxies are computationally expensive, so a more tractable approach is to use less expensive $N$-body simulations that have been populated with galaxies.  The can be accomplished through a technique that matches galaxies to the simulated structure of dark matter, for example, through a halo occupation distribution \citep[HOD, e.g.,][]{2000MNRAS.318.1144P, 2001ApJ...546...20S, 2002ApJ...575..587B, 2005ApJ...633..791Z}.  

Under its simplest assumptions, an HOD uses halo mass as the sole property that determines whether a halo will host a particular type of galaxy.  The breakdown of this assumption is known as galaxy assembly bias, which asserts that mass alone is insufficient and that additional environmental and assembly factors come into play.  These factors include  formation time \citep{2005MNRAS.363L..66G} and halo concentration \citep{2006ApJ...652...71W}.  Modern HOD implementations often provide flexibility to account for assembly bias \citep[e.g.,][]{2016MNRAS.460.2552H, 2018ascl.soft12011Y, 2019arXiv190811448B}.  

Machine learning (ML) offers a number of methods that can find and extract information from complex spatial patterns imprinted on the 3D distribution of galaxies.  ML, therefore, is an enticing approach for inferring cosmological models in spite of myriad complicating effects. One promising class of tools for this task are Convolutional Neural Networks  \citep[CNNs, e.g.~][]{fukushima1982neocognitron, lecun1999object, NIPS2012_4824, DBLP:journals/corr/SimonyanZ14a}, which are often used in image recognition tasks.  CNNs employ many hidden layers to extract image features such as edges, shapes, and textures.  Typically, CNNs pair layers of convolution and pooling to extract meaningful features from the input images, followed by deep fully connected layers to output an image class or numerical label.  Because these deep networks learn the filters necessary to extract meaningful information from the input images, they require very little image preprocessing.  See \cite{2014arXiv1404.7828S} for a review of deep neural networks.

CNNs are traditionally applied to 2D images, which may be monochromatic or represented in several color bands.  2D CNNs can extract information from non-gaussianities in simulated convergence maps, remarkably improving cosmological constraints over a more standard statistical approach \citep[e.g.,][]{2017arXiv170705167S, 2018PhRvD..97j3515G,  2019arXiv190203663R, 2019NatAs...3...93R}, and recent work has extended this to put cosmological constraints on observations using CNNs \citep{2019arXiv190603156F}.

However, the application of CNNs is not limited to flat Euclidean images \citep[e.g.][]{2019A&C....27..130P}, nor is it limited to two dimensions.  The algorithm can be extended to three dimensions, where the third dimension may be, for example, temporal \citep[e.g., video input, as in][]{ji20133d} or spatial \citep[e.g., a data cube, as in][]{DBLP:journals/corr/KamnitsasLNSKMR16}.  \cite{2017arXiv171102033R} employed the first cosmological application of a 3D CNN, showing that the tool can infer the underlying cosmological parameters from a simulated 3D dark matter distribution.  

We present an application of 3D CNNs to learn cosmological parameters from simulated galaxy maps. 
\new{Our hybrid deep learning architecture learns directly from the calculated 2D power spectrum and simultaneously harnesses non-Gaussianities by also learning directly from the raw 3D distribution of galaxies.}  In Section \ref{sec:mocks}, we describe our mock observations: the suite of cosmological simulations (\ref{sec:abacus}), the range of HODs applied to these simulations (\ref{sec:hod}), the training and validation mock observations (\ref{sec:trainset}), and the carefully constructed and independent test mock observations at the \planck cosmology (\ref{sec:test}).  We describe our trio of deep learning architectures, including the hybrid method, in Section \ref{sec:cnn}.  We present our results in Section \ref{sec:results} and a discussion and conclusions in Section \ref{sec:conclusion}.  Appendix \ref{sec:lifecycle} is more pedagogical in nature; it describes how the range of model predictions evolves with training and suggests new tests for assessing a model's fit. 

\clearpage

\section{Methods: Mock Observations}
\label{sec:mocks}

We use the \abacus suite of simulations\footnote{\url{https://lgarrison.github.io/AbacusCosmos/}} \citep{2018ApJS..236...43G, 2019MNRAS.485.3370G} to create three data sets:  a training set, a validation set, and a testing set.  The training set is used to fit the machine learning model; it spans a range of CDM cosmologies and is populated with galaxies in a way to mimic a variety of galaxy formation models.  The validation set is used to assess how well the machine learning model has fit; it  also spans a range of cosmological parameters and galaxy formation models.  The testing set is independent of both the training and validation sets; it is at the \planck fiducial cosmology \citep{2015arXiv150201589P}, built from simulations with initial conditions not used in the training or validation data sets, and populated with galaxies using HODs not used in the training or testing data sets.  The creation of the three data sets are described in the following subsections.

\subsection{\abacus Simulations}
\label{sec:abacus}

The \abacus simulations are a suite of publicly available $N$-body simulations.  The suite includes the \abacus \texttt{1100box} simulations, a sample of large-volume $N$-body simulations at a variety of cosmologies, as well as the \texttt{1100box} \planck simulations, a sample of simulations with cosmological parameters consistent with the \planck fiducial cosmology.

The \abacus \texttt{1100box} simulations are used to create the training and validation sets.  This suite of simulations comprises 40 simulations at a variety of cosmologies that differ for six cosmological parameters:  $\Omega_{CDM}\,h^2$, $\Omega_b\,h^2$, \sig, \Ho, $w_0$, and $n_s$.  The cosmologies for this suite of simulations were selected by a Latin hypercube algorithm, and are centered on the \planck fiducial cosmology \citep{2015arXiv150201589P}.  Each simulation has side length $1100\Mpch$ and particle mass $4\times10^{10}\Msolarh$.  The suite of 40 simulations are phase-matched.

While the \abacus \texttt{1100box} simulations are used to create the training and validation sets, the \abacus \planck simulations are used to create the testing set.  These 20 simulations have cosmological parameters consistent with \cite{2015arXiv150201589P}:  $\Omega_b \, h^2=0.02222$, $\Omega_m\,h^2=0.14212$, $w_0=-1$, $n_s=0.9652$, $\sigma_8=0.830$, $H_0=67.26$, $N_\mathrm{eff}=3.04$.  They have identical side length ($1100\Mpch$) and particle mass ($4\times10^{10}\Msolarh$) to the \texttt{1100box} suite of simulations, but each uses unique initial conditions and none are phase-matched to the \texttt{1100box} simulations.    See \cite{2018ApJS..236...43G} for more details about the \abacus suite of simulations.

\subsection{Halo Occupation Distribution}
\label{sec:hod}

A halo occupation distribution (HOD) is a way to populate dark matter halos with galaxies.  In their most basic form, HODs are probabilisitic models that assume that halo mass is the sole halo property governing the halo-galaxy connection \citep{2002ApJ...575..587B}.  A standard HOD models the probability of a halo hosting a central galaxy, $\overline{n}_\mathrm{central}$, and the mean number of satellites, $\overline{n}_\mathrm{satellite}$, as a function of a single halo property, the mass $M$. The standard HOD by \cite{2007ApJ...659....1Z} gives the mean number of central and satellite galaxies as

\begin{equation}
\begin{split}
	\overline{n}_\mathrm{central} & = \frac{1}{2} \mathrm{erfc}\left[ \frac{\ln(M_\mathrm{cut}/M)}{\sqrt{2}\sigma} \right]\\[3ex]
	\overline{n}_\mathrm{satellite} & = \left[ \frac{M-\kappa M_\mathrm{cut}}{M_1} \right] ^\alpha \overline{n}_\mathrm{central},
\end{split}
\label{eq:HOD}
\end{equation}
where $M_\mathrm{cut}$ sets the halo mass scale for central galaxies, $\sigma$ sets the width of the error function of $\overline{n}_\mathrm{central}$, $M_1$ sets the mass scale for satellite galaxies, $\alpha$ sets the slope of the power law, and $\kappa M_\mathrm{cut}$ sets the limit below which a halo cannot host a satellite galaxy. $M$ denotes the halo mass, and we use the virial mass definition $M_{vir}$. The actual number of central galaxies in a halo follows the Bernoulli distribution with the mean set to $\overline{n}_\mathrm{central}$, whereas the number of satellite galaxies follows the Poisson distributions with the mean set to $\overline{n}_\mathrm{satellite}$.

While this standard HOD populates halos probabilistically according to halo mass, recent variations of the HOD incorporate more flexibility in modeling.  These flexible HODs allow additional halo properties \textemdash{} beyond the halo mass \textemdash{} to inform galaxy occupation \citep[e.g.,][]{2016MNRAS.460.2552H, 2018ascl.soft12011Y}.
The HOD implemented here is one such flexible model; it uses the publicly available \code{GRAND-HOD} package\footnote{\url{https://github.com/SandyYuan/GRAND-HOD}}.  This HOD implementation introduces a series of extensions to the standard HOD, including flexibility in the distribution of satellite galaxies within the halo, velocity distribution of the galaxies, and galaxy assembly bias.  To add this flexibility, we invoke two extensions: the satellite distribution parameter, $s$, and the galaxy assembly bias parameter, $A$. The satellite distribution parameter allows for a flexible radial distribution of satellite galaxies within a dark matter halo, and the galaxy assembly bias parameter allows for a secondary HOD dependence on halo concentration. For complete information about \code{GRAND-HOD} and its HOD extensions, see \cite{2018MNRAS.478.2019Y}.  

Fifteen sets of HOD model parameters are generated for each \abacus simulation box, and 31 are generated for each \planck box.  For each simulation box, a baseline HOD model is selected as a function of cosmology; these baseline models vary only in $M_\mathrm{cut}$ and $M_1$, and  baseline values of all the other HOD parameters remain the same. This ensures that the combined effect of perturbing the cosmology and HOD is mild. This is done because, despite the fact that the cosmological parameters of each simulation are only perturbed by a few percent, coupling these cosmological changes with perturbations to the HOD can lead to drastic changes to the mock catalogs and the clustering statistics.  To minimize these effects, instead of populating galaxies according to HOD parameters in an ellipse aligned with the default parameter basis, we populate according to parameter in an ellipse defined over a custom parameter basis.

\begin{figure*}[]
	\begin{tabular}{c c c}
		
		\includegraphics[width=0.3\textwidth]{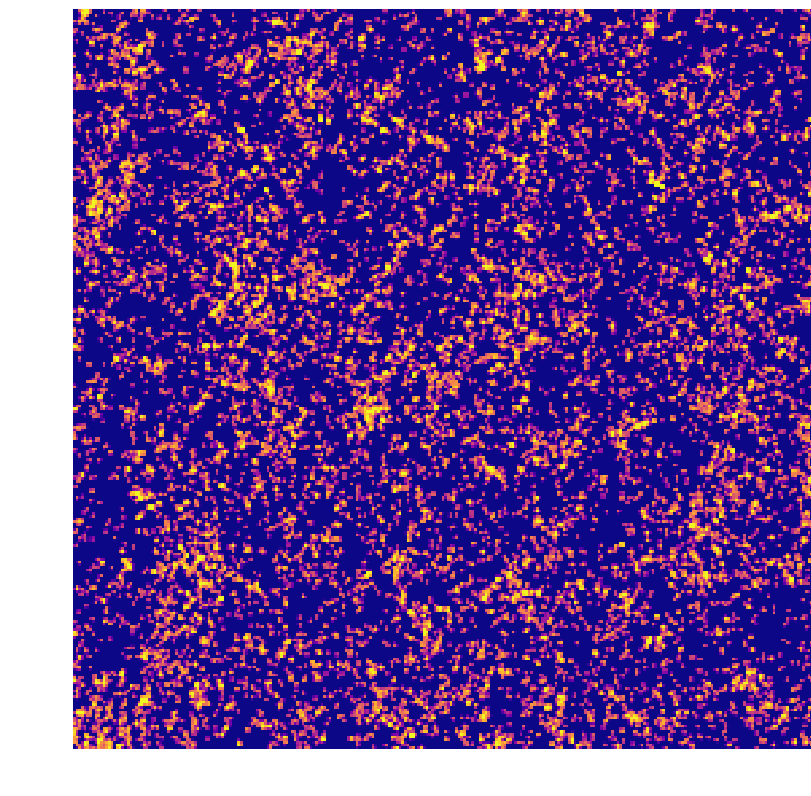} &  \includegraphics[width=0.3\textwidth]{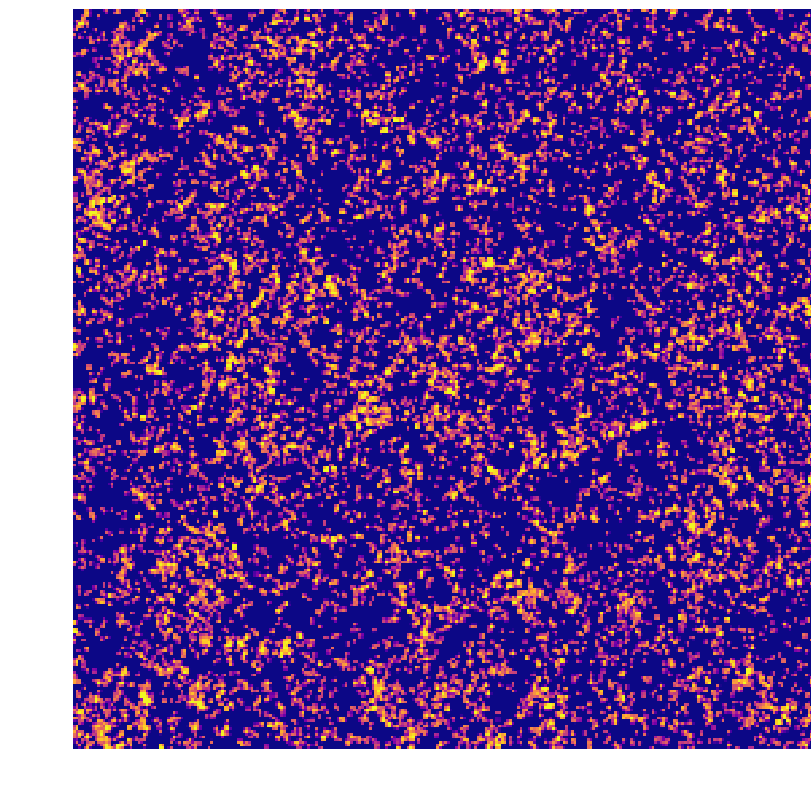} & \includegraphics[width=0.3\textwidth]{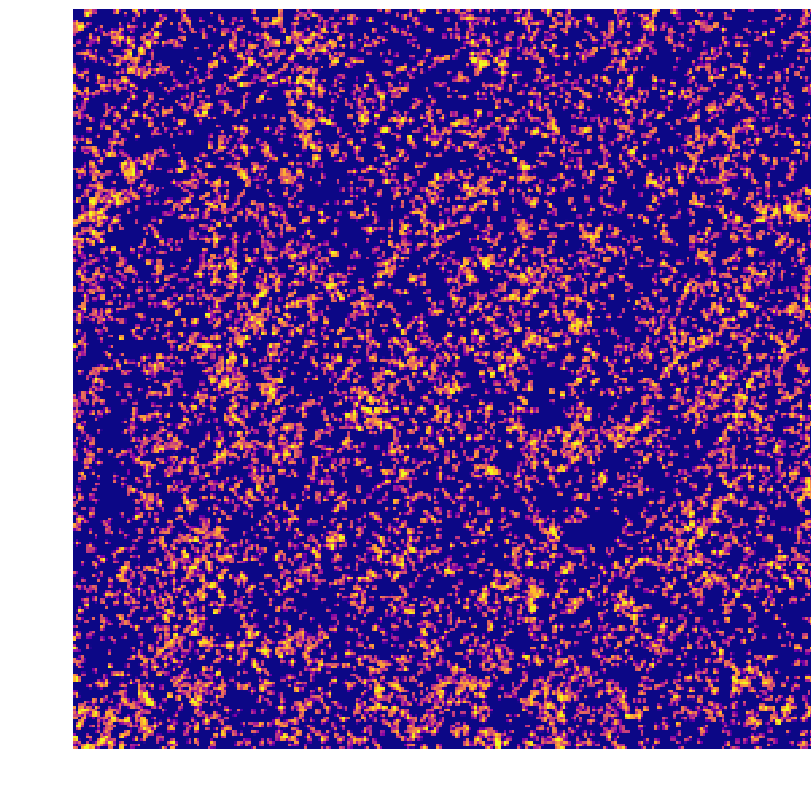}\\[-3ex]
		 $\sig=0.92$,  $\om= 0.28$ & $\sig=0.83 $,  $\om= 0.32$ & $\sig=0.71 $,  $\om= 0.34$ \\[5ex]
		\includegraphics[width=0.3\textwidth]{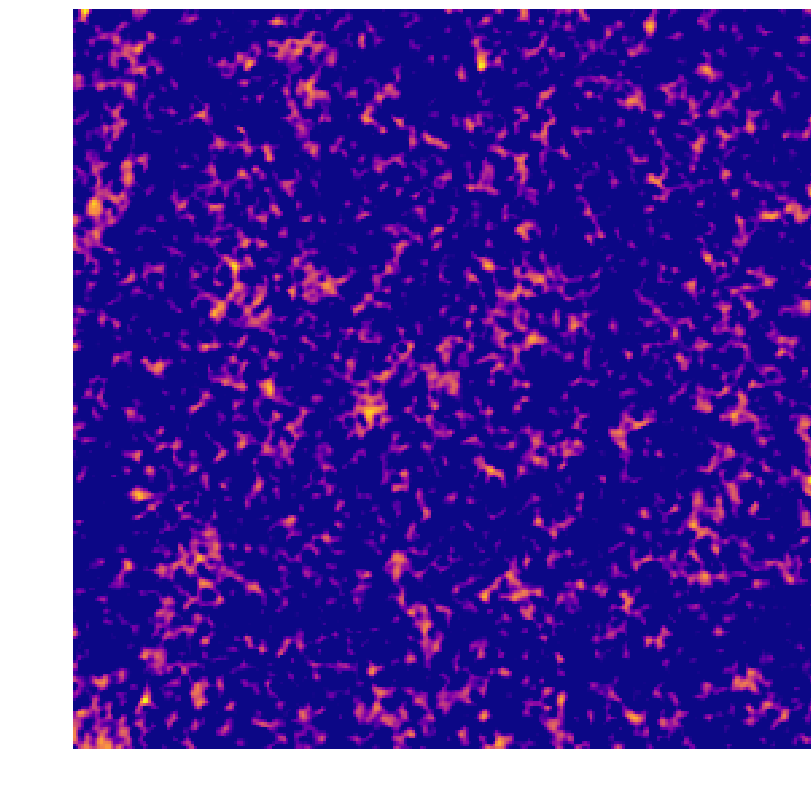} & \includegraphics[width=0.3\textwidth]{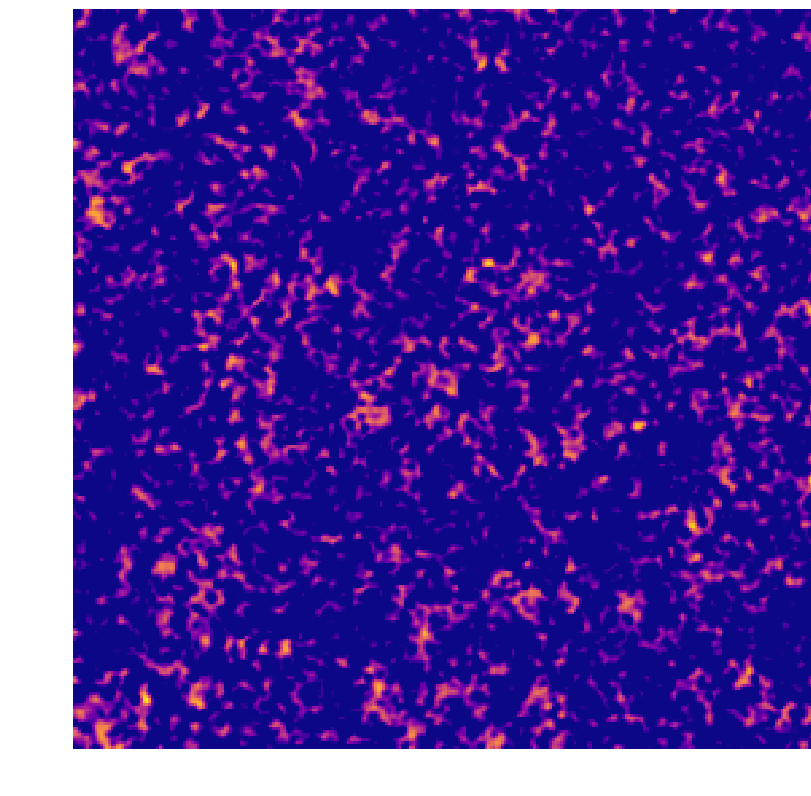} & \includegraphics[width=0.3\textwidth]{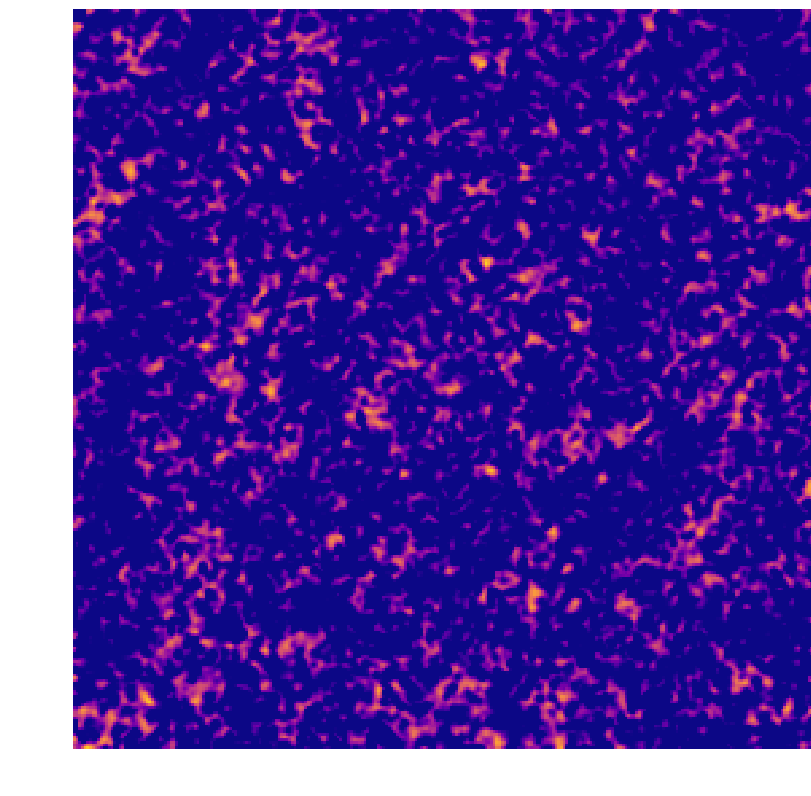}\\[-3ex]
		 $\sig=0.92$,  $\om= 0.28$ & $\sig=0.83 $,  $\om= 0.32$ & $\sig=0.71 $,  $\om= 0.34$ \\[5ex]
		
	\end{tabular}

	\caption[]{Top:  A sample of train input images.  Shown is a two-dimensional projection of the three-dimensional image (or ``slab'').	  The train, validate, and test samples include a number of choices designed to reduce the likelihood of giving the machine learning model an unfair advantage:  we employ a zero-point shift to minimize learning from images with correlated structure, we use random HODs and seeds to allow for uncertainties in galaxy formation physics, we use axial flips of the slabs to augment the data, and we use unique portions of the simulation and unique HODs in the validation set to provide a way to test that the model does not rely on the particulars of the structure or HOD.  To highlight the differences in the images that are strictly due to cosmology and HOD, the zero-point shift has been omitted for these images. Bottom:  The same images as above, smoothed with a Gaussian filter ($\sigma=1\,\mathrm{pixel}$) to  emphasize the differences between images that are due to cosmological models.}
       	\label{fig:slabs}	
\end{figure*}

For the \planck cosmology, the HOD parameters are chosen in reference to the parameter ranges in \cite{2015ApJ...810...35K}:  $\log_{10}(M_\mathrm{cut}/\Msolarh)=13.35$, $\log_{10}(M_1/\Msolarh)=13.8$, $\sigma=0.85$, $\alpha=1$, $\kappa=1$, $s=0$,  and $A=0$.   However, we modify two baseline HOD parameter values \textemdash{} $M_\mathrm{cut}$ and $M_1$  \textemdash{} for the non-\planck simulations. We set the baseline value of $M_\mathrm{cut}$ in each cosmology box such that the projected 2-point correlation function $w_p(5-10\rm{Mpc})$ of all the halos with $M > M_{\mathrm{cut}}$ is equal to the $w_p(5-10\rm{Mpc})$ of the centrals in the baseline HOD at Planck cosmology, where $w_p(5-10\rm{Mpc})$ is defined as
\begin{equation}
    w_p(5-10\textrm{Mpc}) = \int_{5\textrm{Mpc}}^{10\textrm{Mpc}} w_p d (r_\perp).
\end{equation}
This effectively holds the baseline $w_p$ of the centrals approximately constant across all the cosmology boxes. Then $M_1$ is selected such that the baseline satellite-central fraction in each cosmology box is the same as that of the baseline HOD in Planck cosmology.

For each \texttt{1100box}, seven additional pairs of model parameters uniformly sample the parameter space within $5\%$ of the baseline HOD (15 additional pairs for each \planck box). For HOD parameters $s$ and $A$, whose baseline parameters are 0, we draw uniform samples between $-0.05$ and $0.05$. The two HODs of each pair are symmetrically offset across the baseline HOD. Excluding the baseline HOD, fourteen unique HODs are generated for \textit{each} \abacus \texttt{1100box} simulation, and 30 unique HODs are generated for each \planck simulation. Four random seeds are used to populate the simulations with realizations of galaxies according to the HOD; this results in four unique galaxy catalogs for each HOD.  The details of how these are used are described in the next section. For complete information about the HOD implementation, see \cite{2019arXiv190705909Y}.

\subsection{Training \& Validation Sets}
\label{sec:trainset}

The training sample of mock observations (for training the deep learning models) and validation sample of mock observations (for assessing when the models have sufficiently fit) are created from the \abacus suite of \texttt{1100box} simulations.  

\abacus includes 40 simulated cosmologies, and for each of these, we select a random distance along the $x$ and $y$ axes to become the new 0-point of the box ($z=0$, along the line of sight direction, which includes redshift space distortion, remains unchanged).  Because the \texttt{1100box} simulations all have the same initial conditions, this random reshuffling minimizes the chances of our model learning about correlated structure across simulations.\footnote{Simulations with matched initial conditions will produce portions of the cosmic web with, for example, a unique or unusual fingerprint of filamentary structure.  The evolutionary stage of a particular structure is highly dependent on the simulation's \sig and other cosmological parameters.  Because CNNs are particularly adept at pattern finding, care must be taken to prevent a CNN from learning to identify some unique structure \textemdash{} especially one which is particular to a suite of simulations and the initial conditions of those simulations \textemdash{} and infer cosmological parameters from its details.  This is not an approach that will generalize to real observations, and can give overly optimistic results.}  The mock observations of the training set are built from the portion of the box with $220\Mpch \leq z < 1100\Mpch$, while the validation set is built from the structure in the range $0\Mpch \leq z < 220 \Mpch$.  By completely excluding this portion of the simulation from the training set, we can test and ensure that the machine learning model does not rely on its ability to identify or memorize large-scale structure correlations stemming from the matched initial conditions.

The box is divided into 20 non-overlapping slabs, which are $550\Mpch$ in the $x$ and $y$ directions and $220\Mpch$ along the line of sight $z$ direction.  Halo catalogs generated by the \code{ROCKSTAR} halo finder \citep{2012ascl.soft10008B} become the basis for four mock observations per slab.

For each slab, we select and apply one HOD from the 15 that are available.  Eleven of the HODs are reused as necessary in the 16 training slabs.  The remaining four HODs are reserved exclusively for the four validation slabs.  By setting aside four HODs for the validation set, the validation set is populated with galaxies in a way that is unique from the observations used for training, and we can ensure that the ML model results are not dependent on memorization or previous knowledge of the details of the HOD.

For each of the four random HOD seeds, the slabs are populated with galaxies.  These training slabs vary in the number of galaxies, ranging from $\sim17000$ to $\sim46000$ galaxies per slab, 
with the number of galaxies correlating weakly with the underlying cosmology.  To prevent the CNN from learning correlations between cosmological parameters and the number of galaxies in the mock observation, we randomly subselect the galaxy population so that all observations have $15000$ galaxies.

The selected galaxies are binned into a $275\times275\times55$, three-dimensional, single-color image.  Galaxies are assigned to voxels using a triangular shaped cloud (TSC) and $2 \times 2 \times 5\Mpch$ voxels.  Projected galaxy densities for three sample cosmologies are shown in Figure \ref{fig:slabs}.

Because the machine learning model described in Section \ref{sec:cnn} is not invariant under mirroring of images, we augment our data by applying an axial flip along the $x$- and/or $y$-directions to three of the four slabs.  For each of these three mirror images, we use a new random seed for the HOD and uniquely subselect to 15000 galaxies.  

\begin{figure}[]
	\begin{centering}
		\includegraphics[width=0.5\textwidth]{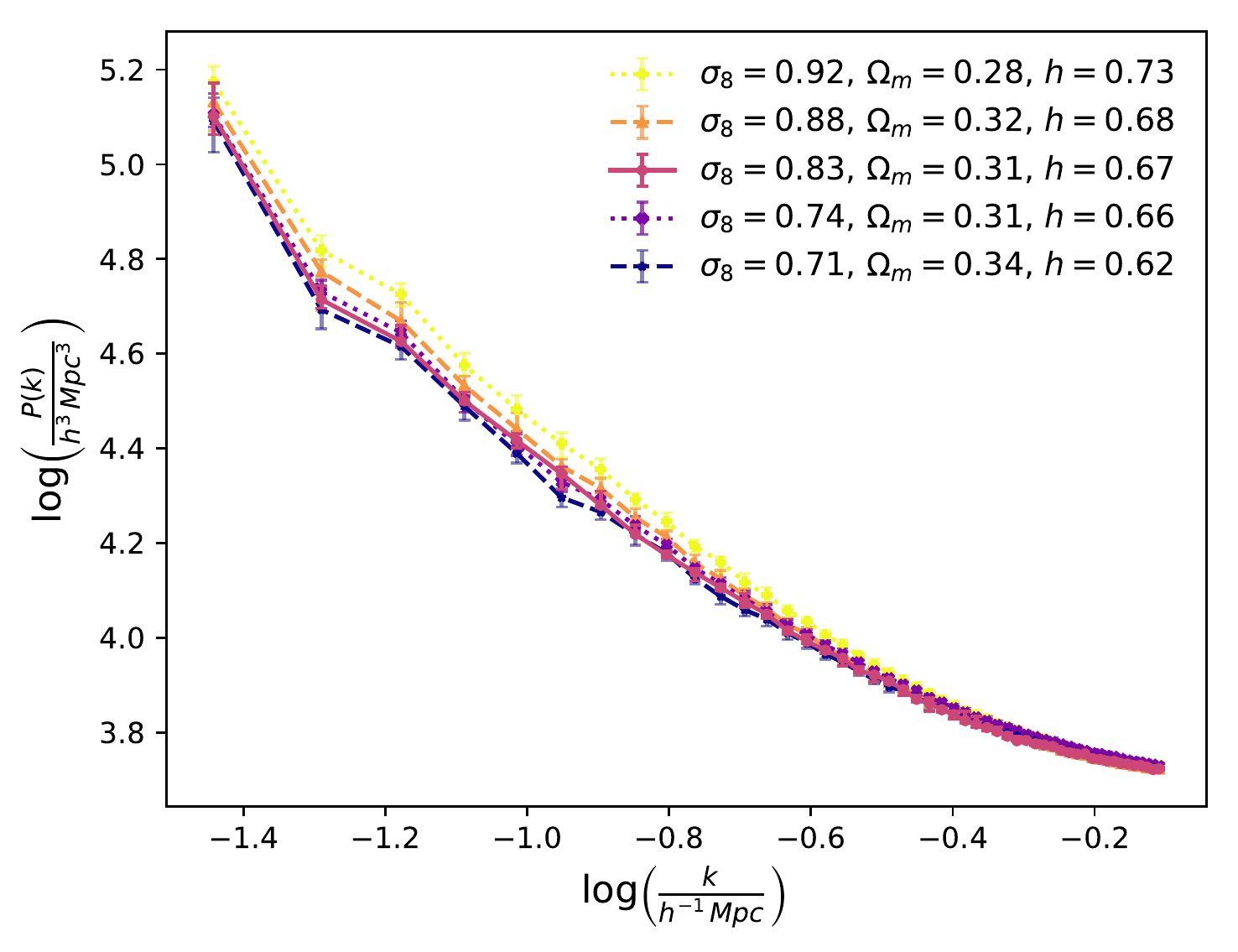}\\
	\end{centering}
	\caption[]{Mean galaxy power spectra, $P(k)$, for 4 of the 40 training cosmologies (yellow, orange, purple, and blue) as well as for the \planck test cosmology (pink).  Points indicate the mean power, while error bars show the middle 68\% of the mock observations.  The ``Vector Features'' input, shown in Figure \ref{fig:architecture}, is a single realization of this power spectrum; for each mock observation, the power spectrum is calculated directly from a single 3D mock galaxy observation.}
       	\label{fig:ps}	
\end{figure}

The power spectrum of the galaxy density field is computed for each slab.  To perform this calculation, we pad the galaxy density field with zeros to double the image size in each direction to account for the lost periodic boundary conditions, Fourier transform the resulting $550\times550\times110$ image, and convert the result to a power spectrum in physical units.  This 3-dimensional power spectrum is next de-convolved to account for the TSC-aliased window function \citep[as in, e.g.,][]{2010PhDT.........4J}, and summarized as a 1-dimensional power spectrum by averaging the power in binned spherical annuli.  Due to the anisotropic nature of the slab and voxel dimensions, the most conservative choices for minimum and maximum $k$ values are selected. These are set by the shortest box dimension ($220\Mpch$) and the Nyquist frequency of the largest pixel dimension ($5\Mpch$), respectively.  Power spectra for a sample of galaxy catalogs are shown in Figure \ref{fig:ps}. 

To recap, the method for building mock observations from each of the simulations is as follows:
\begin{itemize}
\itemsep-0.3em 
	\item A random $x$ and $y$ value is selected to be the new 0-point of the box.  $z=0$, along the line of sight direction with redshift space distortion, remains unchanged.  
	\item The box is divided into 20 non-overlapping slabs, each $550 \times 550\times 220\Mpch$.
	\item For each slab: \\[-4 ex]
	\begin{itemize}
		\itemsep-0.3em 
		\item An HOD is selected.  Eleven HODs, some of which are reused as necessary, are used to populate the 16 training slabs with galaxies.  Four unique HODs are reserved exclusively for the four validation slabs.  
		\item 15000 galaxies are randomly selected.  These are binned in $2\times 2\times 5\Mpch$ bins using a TSC.
		\item The previous step is repeated for each of 4 random seeds, incorporating mirror image(s) of the slab.
		\item The power spectrum of the slab is calculated.
	\end{itemize}
\end{itemize}
This method results in 3200 mock observations built from 40 simulations, with 20 slabs per simulation and 4 seeds (with axial flips) per slab.

The 2560 slabs built from the portion of the simulation with $z\geq220\Mpch$ comprise the training set, and are used to train the machine learning model described in Section \ref{sec:cnn}.  The remaining 640 slabs are built from a non-overlapping  portion of the simulation ($z<220\Mpch$).  These make up the validation set and are used to assess the models' fit.

Our creation of the test and validation sets include a number of choices to reduce the likelihood of giving the machine learning model an unfair advantage:  we employ a recentering of the box to minimize learning from images with correlated structure, we use random HODs and seeds to allow for uncertainties in galaxy formation physics, we use axial flips of the slabs to augment the data to account for  rotational invariance, and we use unique portions of the simulation and unique HODs in the validation fold to provide a way to ensure that the model does not rely on the details of the structure or HOD.

\subsection{Planck Testing Set}
\label{sec:test}

The testing sample is built from the \abacus \planck simulations.  The 20 \planck simulations each have initial conditions that are unique from the simulation sample described in Section \ref{sec:trainset}.  Mock observations of the \planck testing set are built using a similar process as described in Section \ref{sec:trainset} with one exception:  the 20 non-overlapping slabs are each populated with galaxies according to 20 unique HODs selected randomly from the 31 HODs available.  Accounting for the axial flips to augment the data, the resulting testing sample is 1600 slabs with associated power spectra. Our testing set is a truly independent sample from the training and validation sets.  Though the cosmologies used in the training and validation sets are near the \planck fiducial cosmology, this exact cosmology is never explicitly used for training or testing.  

\section{Methods:  Machine Learning Models}
\label{sec:cnn}

We assess three machine learning models:  
1.~a standard convolutional neural network (CNN) that learns from the 3D galaxy images to regress cosmological parameters,  
2.~a neural network (NN) that learns from the power spectrum of the galaxy images to regress cosmological parameters, and 
3.~a hybrid CNN (hCNN) that employs a standard CNN but also can take advantage of meaningful summary information \textemdash{} in this case, the galaxy power spectrum \textemdash{} to inject physically meaningful information into the fully connected layers.  These three models are described in detail below.

\subsection{Standard CNN} 

\begin{figure}[]
	\begin{centering}
		\includegraphics[width=0.45\textwidth]{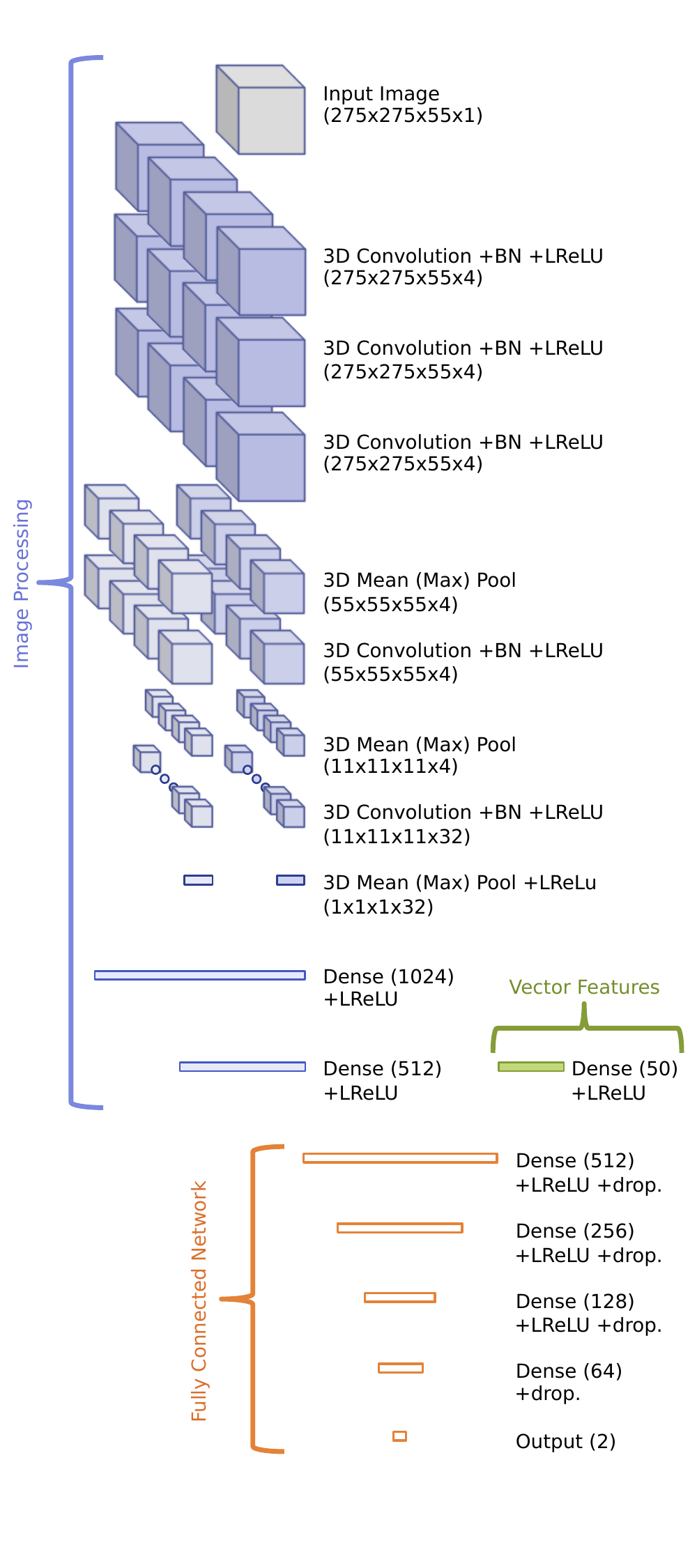}\\
	\end{centering}
	\caption[]{A visual summary of the three ML models.  The neural network (NN) uses a vector input (green) with the fully connected layers for processing (orange).  The standard convolutional neural network (CNN) uses an image input with the image processing layers (blue) plus fully connected layers (orange).  The hybrid CNN (hCNN) joins these by concatenating the vector features with the final layer of the image processing; the result is fed into the fully connected layers. For further details about the NN, CNN, and hCNN, see Section \ref{sec:cnn}.\\}
       	\label{fig:architecture}	
\end{figure}

Convolutional Neural Networks \citep[CNNs, ][]{fukushima1982neocognitron, lecun1999object, NIPS2012_4824} are a class of machine learning algorithms that are commonly used in image recognition tasks.  Over many cycles, called ``epochs,'' the network learns the convolutional filters, weights, and biases necessary to extract meaningful patterns from the input image.  For cosmological applications, CNNs are traditionally applied to monochromatic \citep[e.g.,][]{2018MNRAS.473.3895L, 2018arXiv181007703N, 2019arXiv190205950H} or multiple-color 2D images \citep[e.g.,][]{ 2018arXiv181008211L}.  However, CNNs are not confined to 2D training data; they can also be used on 3D data cubes.  Three-dimensional CNNs became popular for interpreting videos, using time as the third dimension \citep[e.g.,][]{ji20133d}, but recent cosmological applications of this algorithm have applied the technique to 3D data \citep[e.g.,][]{2017arXiv171102033R, 2018arXiv181106533H, 2018arXiv180804728M, 2018arXiv181011030P, 2019MNRAS.484.5771A, 2019MNRAS.482.2861B, 2019arXiv190205965Z, 2019arXiv190810590P}.

CNNs typically use pairs of convolutional filters and pooling layers to extract meaningful patterns from the input image.  These are followed by several fully connected layers. Our standard CNN architecture includes several consecutive fully convolutional layers at the onset and mean and max pooling branches in parallel. It is implemented in Keras \citep{chollet2015} with a Tensorflow \citep{45381} backend, and is shown in Figure \ref{fig:architecture}.  The full architecture is as follows:
	\begin{enumerate}
	\itemsep-0.3em  
		\item $3\times3\times3$ convolution with 4 filters
		\item[]leaky ReLU activation
		\item[]batch normalization
		\item $3\times3\times3$ convolution with 4 filters
		\item[]leaky ReLU activation
		\item[]batch normalization
		\item $3\times3\times3$ convolution with 4 filters
		\item[]leaky ReLU activation
		\item[]batch normalization
		\item Max pooling branch (in parallel with \# \ref{item:meanpool}): \label{item:maxpool}
			\begin{enumerate}
			\itemsep-0.2em  
				\item $5\times5\times1$ max pooling
				\item $3\times3\times3$ convolution with 4 filters
				\item[]leaky ReLU activation
				\item[]batch normalization
				\item $5\times5\times5$ max pooling
				\item $3\times3\times3$ convolution with 32 filters
				\item[]leaky ReLU activation
				\item[]batch normalization
				\item $5\times5\times5$ max pooling, flattened \label{item:maxpooloutput}
			\end{enumerate}
		\item Mean pooling branch (in parallel with \# \ref{item:maxpool}): \label{item:meanpool}
			\begin{enumerate}
			\itemsep-0.2em  
				\item $5\times5\times1$ max pooling
				\item $3\times3\times3$ convolution with 4 filters
				\item[]leaky ReLU activation
				\item[]batch normalization
				\item $5\times5\times5$ max pooling
				\item $3\times3\times3$ convolution with 32 filters
				\item[]leaky ReLU activation
				\item[]batch normalization
				\item $5\times5\times5$ max pooling, flattened \label{item:meanpooloutput}
			\end{enumerate}
		\item Concatenation of the max pool branch output (\ref{item:maxpooloutput}) and mean pool branch output (\ref{item:meanpooloutput}) 
		\item[]leaky ReLU activation
		\item 1024 neurons, fully connected
		\item[]leaky ReLU activation
		\item[] 30\% dropout
		\item 512 neurons, fully connected \label{item:nnstart}
		\item[]leaky ReLU activation
		\item[] 30\% dropout
		\item 512 neurons, fully connected
		\item[] leaky ReLU activation 
		\item[] 30\% dropout
		\item 256 neurons, fully connected
		\item[] leaky ReLU activation
		\item[] 30\% dropout
		\item 128 neurons, fully connected
		\item[]leaky ReLU activation 
		\item[] 30\% dropout
		\item 64 neurons, fully connected
		\item[] linear activation 
		\item[] 30\% dropout
		\item 2 output neurons, one each for \om and \sig \label{item:nnend}
	\end{enumerate}
We use a mean absolute error loss function and the Adam Optimizer \citep{2014arXiv1412.6980K}. In practice, we scale \om and \sig linearly so that the range of training values lies between $-1$ and $1$.  The output predictions are scaled back to physically interpretable values according to the inverse of the same linear scaling.  While this may not be an important detail for these particular cosmological parameters (\sig and \om are of the same order of magnitude), problems can arise when training multiple outputs with significantly different value ranges (e.g. if \Ho in units of $\mathrm{km}\,s^{-1}\,\mathrm{Mpc}^{-1}$ were added as a third output parameter).  Details about the training scheme and learning rate are discussed in Section \ref{sec:training}.

In our model, small-scale feature extraction is performed by several consecutive layers of 3D $3\times3\times3$ convolutional filters.  This feature extraction is followed by aggressive pooling in parallel max and mean pooling branches that each reduce the data cube to 32 neurons.  The outputs of these branches are concatenated and are followed by fully connected layers.  We use a rectified linear unit \citep[ReLU,][]{nair2010rectified} activation function throughout.  The dropout, in which 30\% of neurons are ignored during training, reduces the likelihood of the model overfitting \citep{srivastava2014dropout}.

The model takes a $275\times275\times55$ image as input and learns the filters, weights, and biases necessary to regress two cosmological parameters \textemdash{} the amplitude of matter fluctuations (\sig) and the matter density parameter (\om); each of the two output neurons maps to a cosmological parameter.  

\subsection{Standard NN}
\label{sec:nn}

The standard neural network uses only the fully connected layers, with the power spectrum as the only input, fed into steps \ref{item:nnstart} through \ref{item:nnend} in the above architecture.  It is shown in Figure \ref{fig:architecture}.  The model takes the binned power spectra as input and learns the weights and biases necessary to regress the cosmological parameters  of interest.

\subsubsection{Hybrid CNN}
\label{sec:mcnn}

The hybrid convolutional neural network (hCNN) takes advantage of a standard CNN, but also utilizes information that is known to be important and meaningful.  The power spectrum, which carries cosmological information, is folded in by inserting this information at step \ref{item:nnstart} in the standard CNN architecture.  It should be noted that the use of incorporating physically meaningful parameters into a deep learning technique is not new to this work, and has been used previous in astronomy \citep{2019arXiv190310507D}, though it has not yet been widely adopted.

The hCNN model uses both the $275\times275\times55$ images as well as the binned power spectra as input to learn  \om and \sig.  This architecture is shown in Figure \ref{fig:architecture}.

\begin{figure*}[]
	\begin{tabular}{c c}
		\includegraphics[width=0.5\textwidth]{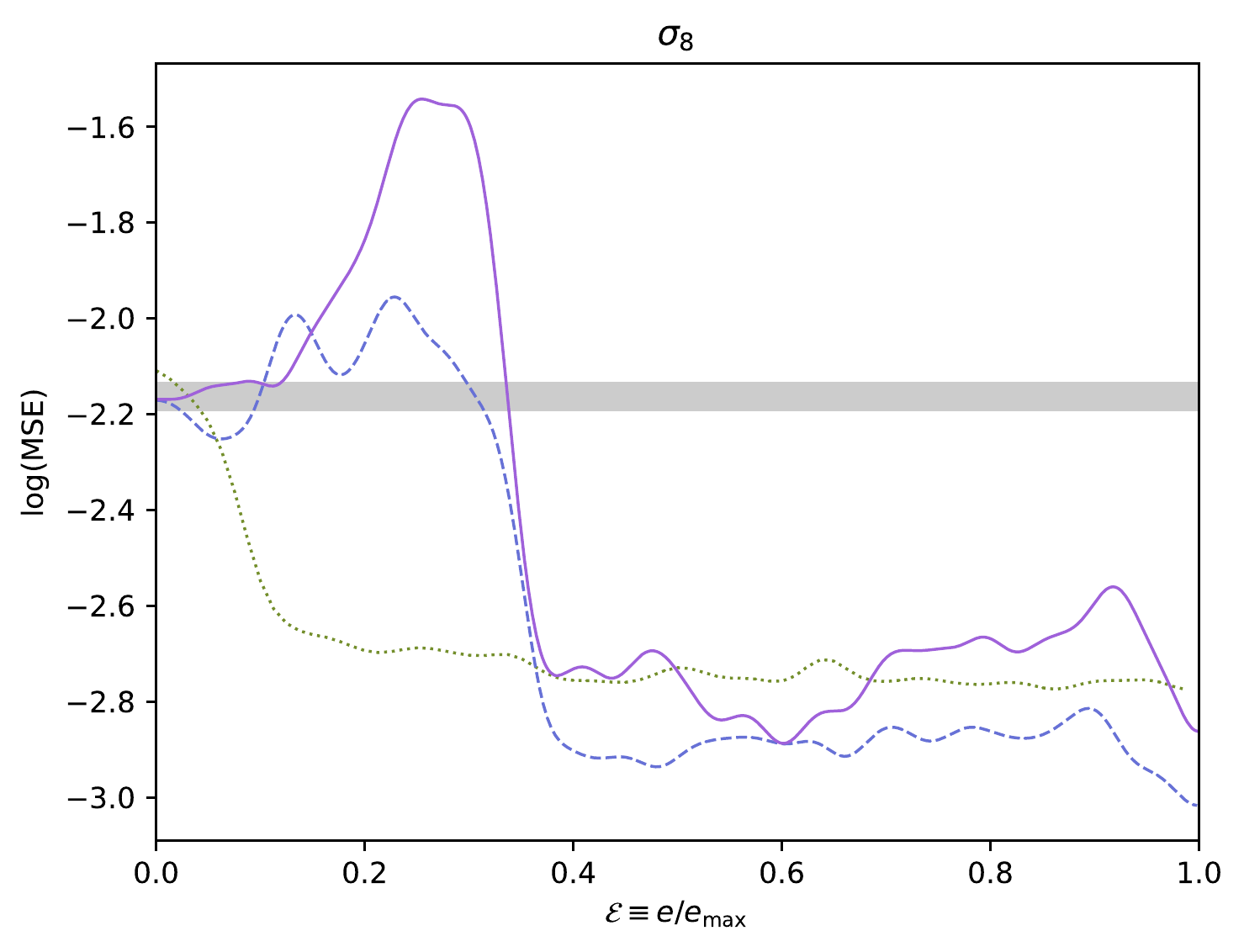} & \includegraphics[width=0.5\textwidth]{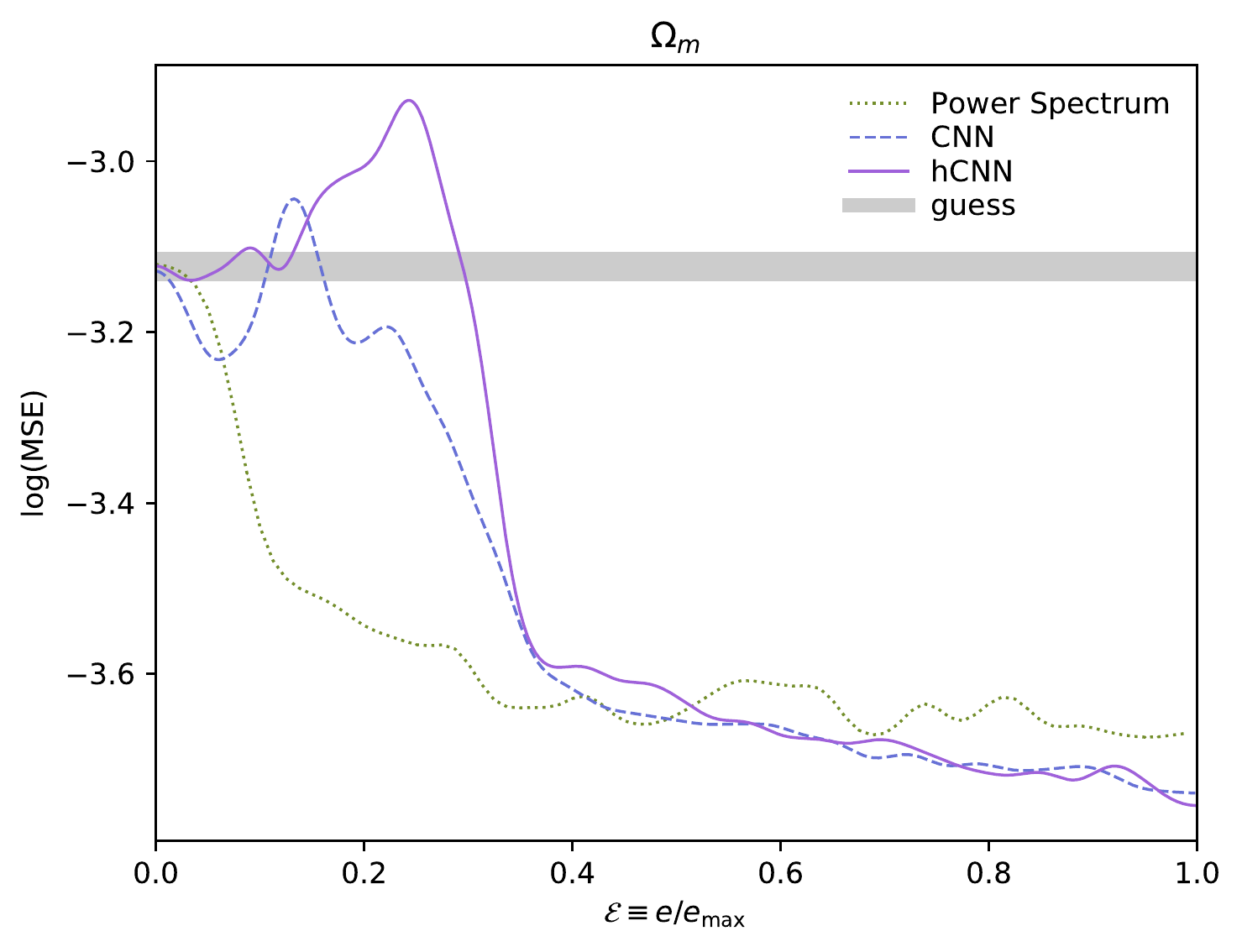} \\
		\end{tabular}
	\caption[]{Mean squared error (MSE) as a function of scaled epoch, $\mathcal{E}$.  While the standard neural network (NN, green dotted) quickly settles to a low error solution, the convolutional neural network (CNN, blue dashed) and hybrid CNN (hCNN, purple solid) have large fluctuations during the initial phase of training ($\mathcal{E}\lesssim0.32$).  Here, the error on the validation set predictions are regularly worse than a guess of the mean value (gray line) for both \sig (left) and \om (right).  The learning rate is decreased at $\mathcal{E}\approx0.32$, and the CNN and hCNN settle into a low-error regime.  
	To remove fluctuations that visually detract from overall trends in error and slope, the curves shown in this figure have been smoothed with a Gaussian filter.}
       	\label{fig:err}	
\end{figure*}

\subsection{Training}
\label{sec:training}

For training the CNN and hCNN, we adopt a two-phase training scheme.  Our training approach  takes advantage of a large step size during the initial phase of training to capture the diversity of cosmologies and HOD models, then transitions to a smaller step size during the second phase of training to improve the fit (see Appendix \ref{sec:lifecycle} for a further discussion of this).  We train for 550 epochs, 175 in the first phase and 375 in the second phase.  The last 50 epochs will be used to select a model that meets criteria more nuanced than simply minimizing the loss function.  It is discussed further in Section \ref{sec:stop}.  Note that the NN, which is less sensitive to the details of training and trains significantly faster than models with convolutional layers, is trained for 800 epochs according to the details of phase one, described below.

We use the Adam Optimizer \citep{2014arXiv1412.6980K}, which has a step size that varies as a function of epoch according to
\begin{equation}
	\alpha(t) = \alpha_0 \frac{\sqrt{1-\beta_2^t}}{1-\beta_1^t},
	\label{eq:stepsize}
\end{equation}
where $\alpha$ is the step size, $t$ denotes a time step or epoch, $\alpha_0$ is the \textit{initial} step size\footnote{The initial step size is denoted, simply, ``learning rate'' in the keras documentation.}, and parameters $\beta_1$ and $\beta_2$ control the step size at each epoch.  We adopt the default values of $\beta_1=0.9$ and $\beta_2=0.999$.

Phase one of training is 175 epochs with an initial step size of $\alpha_0=1.0\times10^{-5}$.  We find that this first phase, with its larger initial step size, is necessary for the models to learn the diversity of cosmologies.  Smaller learning rates tend to produce models with predictions that cluster near the mean values for \sig and \om, while larger learning rates tend to produce models that fluctuate wildly in bias or overfit the training data.  Near epoch 175, we find evidence in the CNN and hCNN that the learning rate is too large. This is characterized by swings in the tendency to over- or underpredict the validation set, and can be seen in the large, fluctuating mean squared error (MSE) shown in Figure \ref{fig:err}.  The MSE is plotted as a function of scaled epoch, $\mathcal{E}$, defined as epoch divided by the maximum number of training epochs.    

We  adopt the model at epoch 175 as a pre-trained model and transition to a second phase of training with a lower learning rate.  Phase two of training is an additional 375 epochs with an initial step size of $\alpha_0=0.2\times10^{-5}$.  For clarity, we refer to the first training epoch of phase two as ``epoch 176'' for the remainder of this work.  However, for the purposes of Equation \ref{eq:stepsize} only, $t$ is reset to $0$.   Figure \ref{fig:err} shows the effect of decreasing the learning rate:  at $\mathcal{E}\approx 0.32$, the mean squared error decreases dramatically as the model settles into a stable fit that describes the validation data.

Overfitting is defined as the tendency of the model to produce excellent predictions on the testing set but to fail on the validation set. (The term ``overfit'' is occasionally used to describe a deep learning method identifying features in a cosmological simulation that do not describe actual observations, but we use the term in the more traditional sense.) Two changes to the learning scheme tend to result in an overfit model:  first, an increased learning rate and second, the use of max pooling only via  eliminating the mean pooling branch.  When the model is overfit, the validation set dramatically biases toward the mean, despite the fact that the training data are well-described even at extreme values of \sig and \om.  

We caution, however, that we have not explored a full grid of hyperparameters for model optimization.  It is likely that the two-phase training scheme could be avoided with carefully selected values of $\beta_1$ and $\beta_2$ to smoothly decrease step size.  Likewise, we have not thoroughly vetted the tendency to overfit by increasing learning rate or removing mean pooling under many hyperparameter combinations.  Such a comprehensive grid search is  expensive and intractable with current computational resources.  Therefore, the effects of learning rate and pooling described in this section should serve as a word of caution for those training other deep models, but should not be overinterpreted.

\section{{Results}}
\label{sec:results}

\begin{figure*}[]
	\begin{tabular}{c c}
		\includegraphics[width=0.5\textwidth]{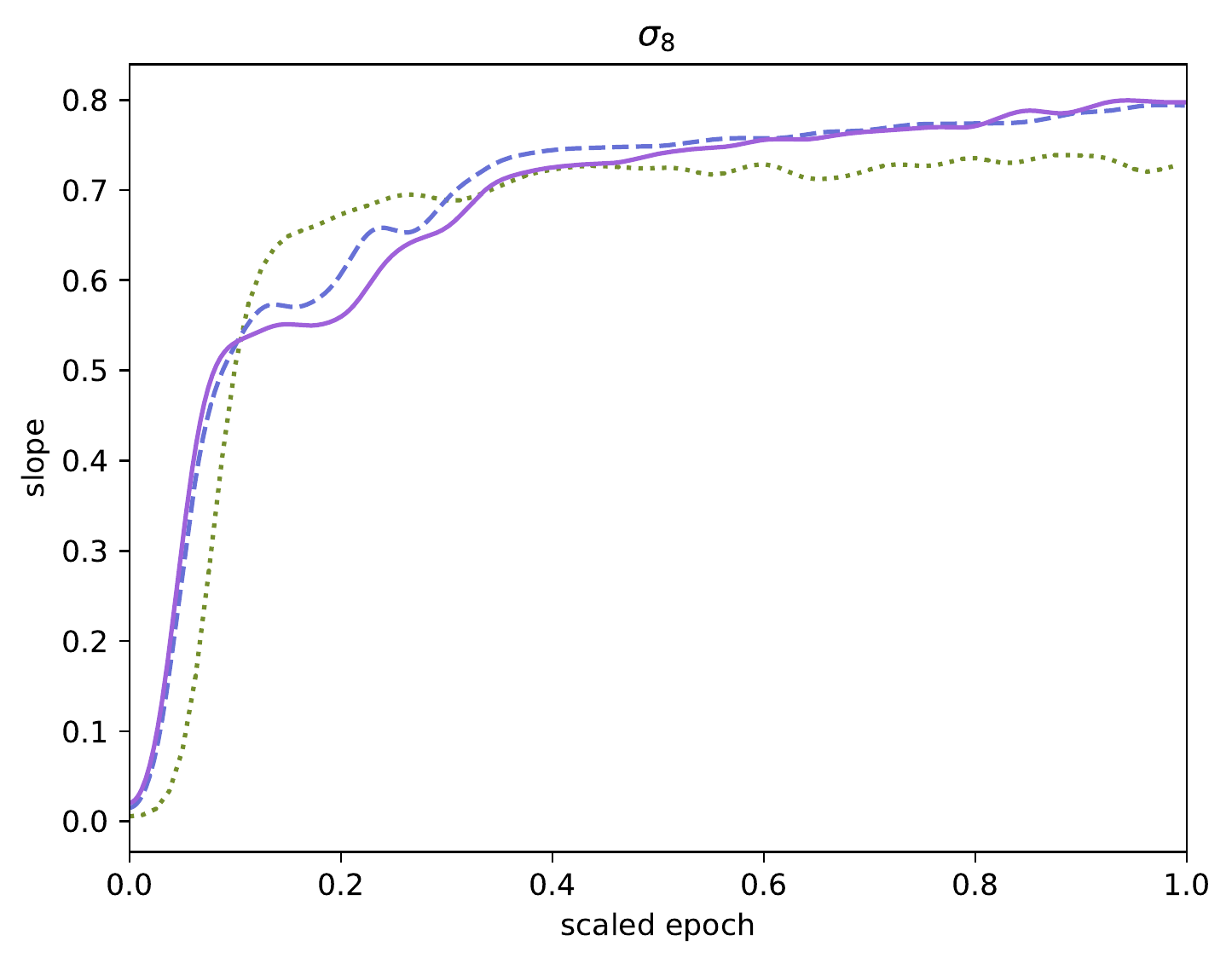} & \includegraphics[width=0.5\textwidth]{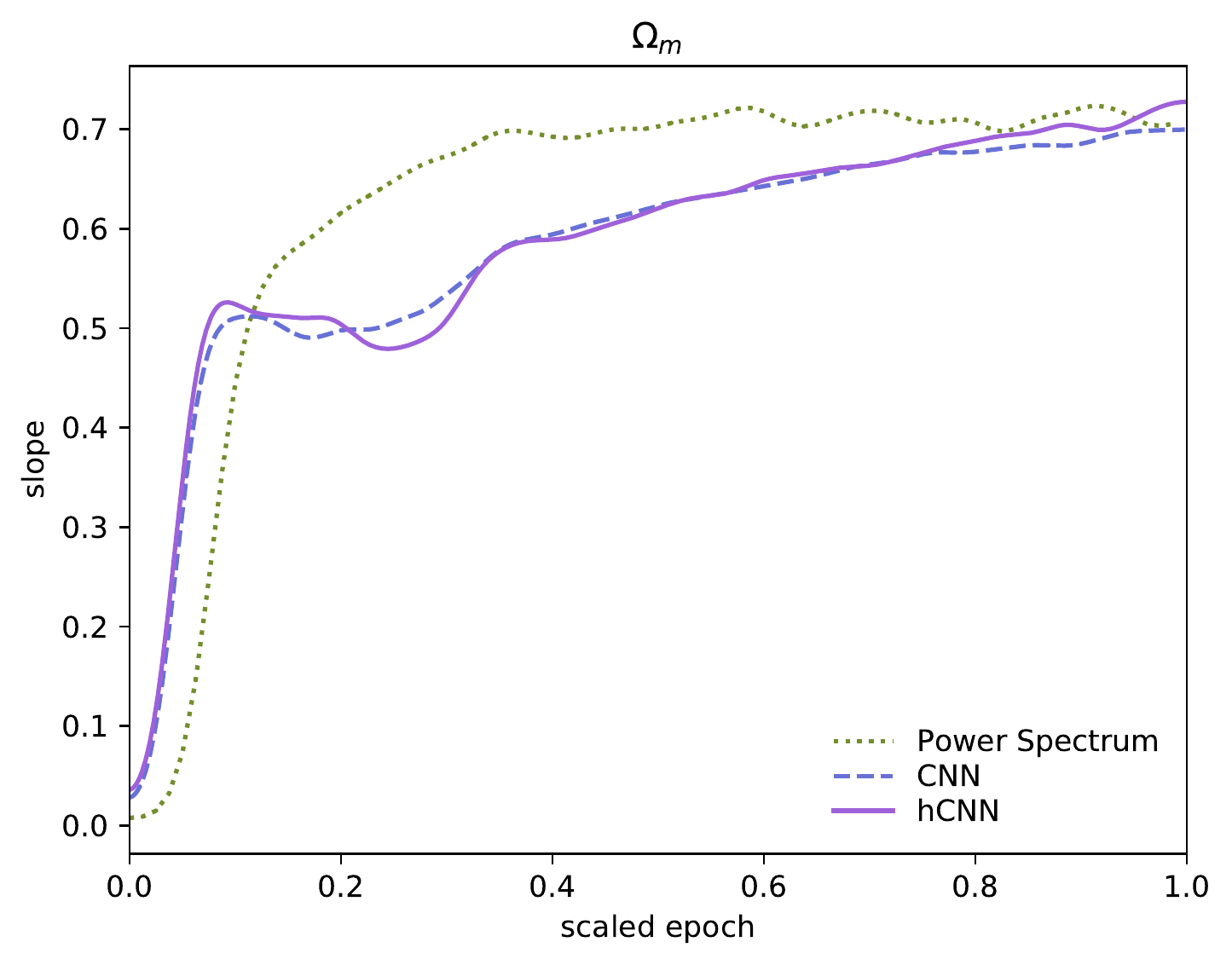}\\ \includegraphics[width=0.5\textwidth]{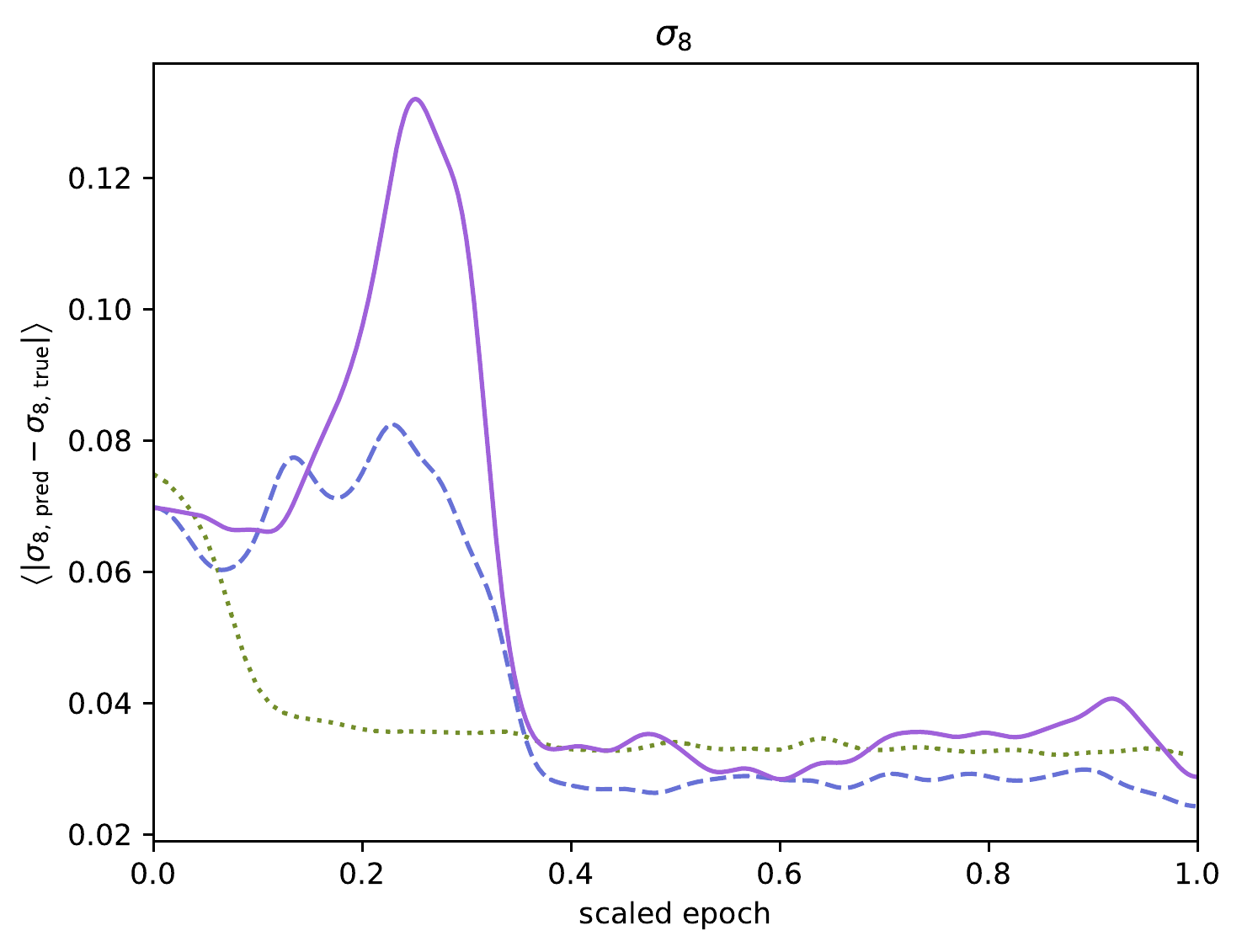} & \includegraphics[width=0.5\textwidth]{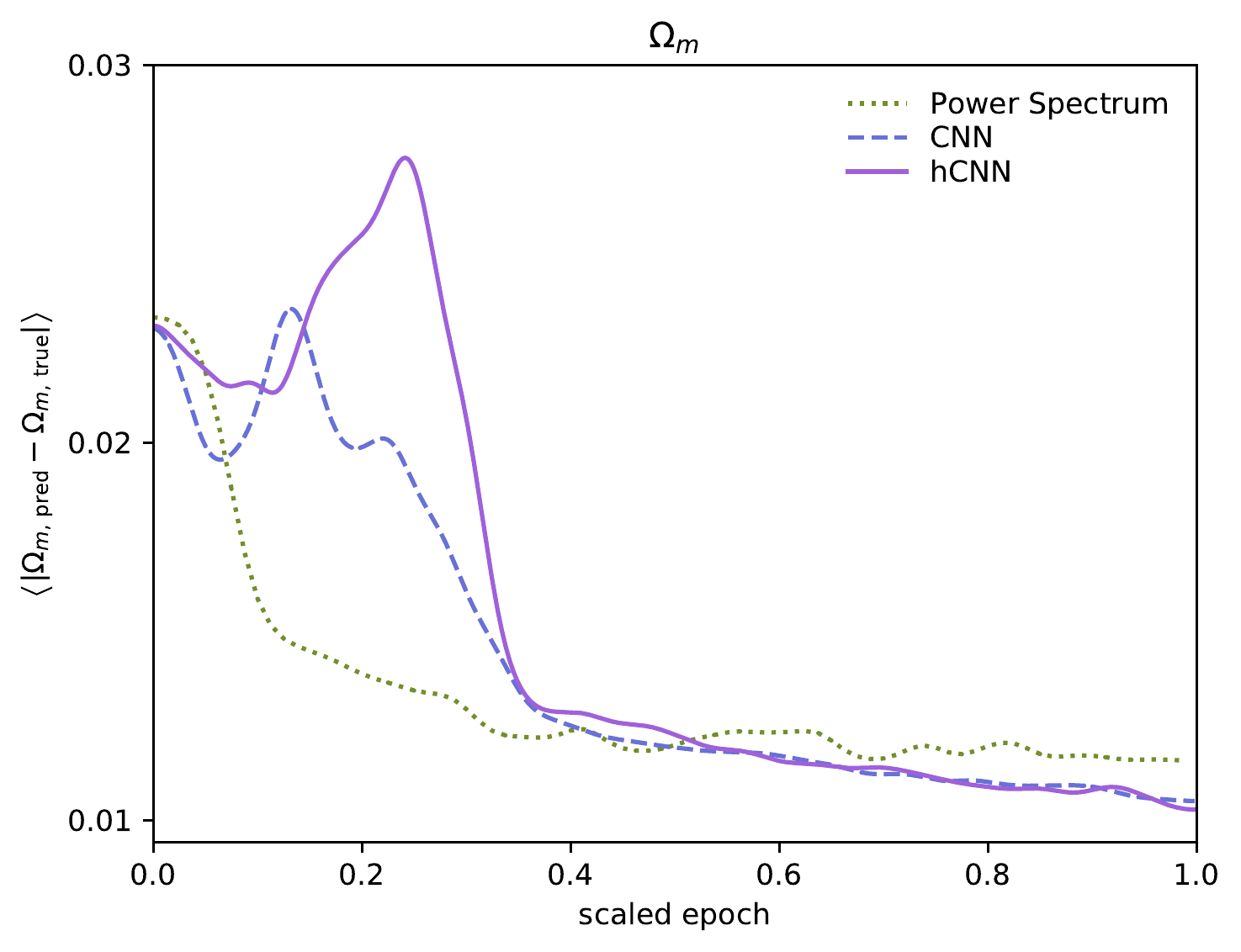}
		\end{tabular}
	\caption[]{Top:  slope of the best fit line as a function of scaled epoch, $\mathcal{E}$.  A slope of 1 indicates that the model captures the full range of \sig and \om, while a slope of 0 is indicative of the model predicting at or near the mean for all data in the validation set.  As the models train, they increase the diversity of predictions.  However, slope never reaches a value of $1$ for any model, indicating that the predictions will bias toward the mean for any mock observation with extreme values of \sig or \om.
	Bottom: prediction bias, b, as a function of scaled epoch, $\mathcal{E}$.  While the standard neural network (NN, green dotted) quickly settles to a solution with low bias, the convolutional neural network (CNN, blue dashed) and hybrid CNN (hCNN, purple solid) have large fluctuations during the initial phase of training ($\mathcal{E}\lesssim0.32$).  The learning rate is decreased at $\mathcal{E}=0.32$, and the CNN and hCNN settle into a low-bias regime.  
	To remove fluctuations that visually detract from overall trends in error and slope, the curves shown in this figure have been smoothed with a Gaussian filter.} 
       	\label{fig:biasandslope}	
\end{figure*}

 Here, we present results from the validation set as a way of assessing the model's fit, both near the median model and also toward extreme values of \sig and \om.  We also present results from the testing set to explore how the technique might generalize into the more realistic case where the cosmological model, galaxy formation details, and initial conditions are not explicitly known.

\subsection{{Validation Set Results}}

We define the prediction bias, $b$, as 
\begin{equation}
	b\equiv \left< \left| x_\mathrm{predicted} - x_\mathrm{true} \right| \right>,
	\label{eq:bias}
\end{equation}
where $<\cdot>$ denotes a mean and $x$ is a placeholder for either \sig or \om. Figure \ref{fig:biasandslope} shows the bias as a function of scaled epoch, $\mathcal{E}$.  During phase two of the training, the CNN and hCNN bias drop significantly, indicating that the lower learning rate is indeed reducing errors in a meaningful way and learning the spatial galaxy patterns that correlate with cosmological parameters.

While MSE and bias both assess the typical offset of the validation set predictions, these statistics alone cannot tell the full story.  It is also important to understand how the model might perform near the edges of the training set.  For this, we assess the slope of a best fit line through the true and predicted values of \sig, and separately, the best fit line through the true and predicted values of \om.  A slope close to $1$ indicates that the model fits well near the extreme values of \sig and \om, while a slope of $0$ is indicative of a model biasing toward the mean.  Overfit models will tend to have a larger MSE and bias coupled with a smaller slope. Figure \ref{fig:biasandslope} shows the slope of this linear best fit line.  We can infer from the value of this fit, $\sim0.7-0.8$ for both \sig and \om, that the model may not predict well for \sig and \om values near the edges of the training data, and will likely bias toward the mean when presented with a cosmological parameter set far from the mean.

\subsection{{Testing Set Results}}

While it is an interesting academic exercise to discuss the results of the validation set, the Universe, unfortunately, gives us one galaxy sample.  This sample  may differ from our training set in cosmological parameters and galaxy formation physics (and most certainly differs in initial conditions!).  If we aim to eventually use a CNN or hCNN to constrain cosmological models from an observed galaxy sample, is imperative to develop tools to assess ML models, going beyond a simple minimization of loss or performance on validation data.  Though the model trains to minimize the mean absolute error, this is not necessarily the most interesting \textemdash{} or the most useful \textemdash{} test statistic for a cosmological analysis of a large galaxy survey. Next, we lay out a technique for selecting a relatively unbiased model.  

\begin{figure*}[]
	\begin{tabular}{c c}
		\includegraphics[width=0.45\textwidth]{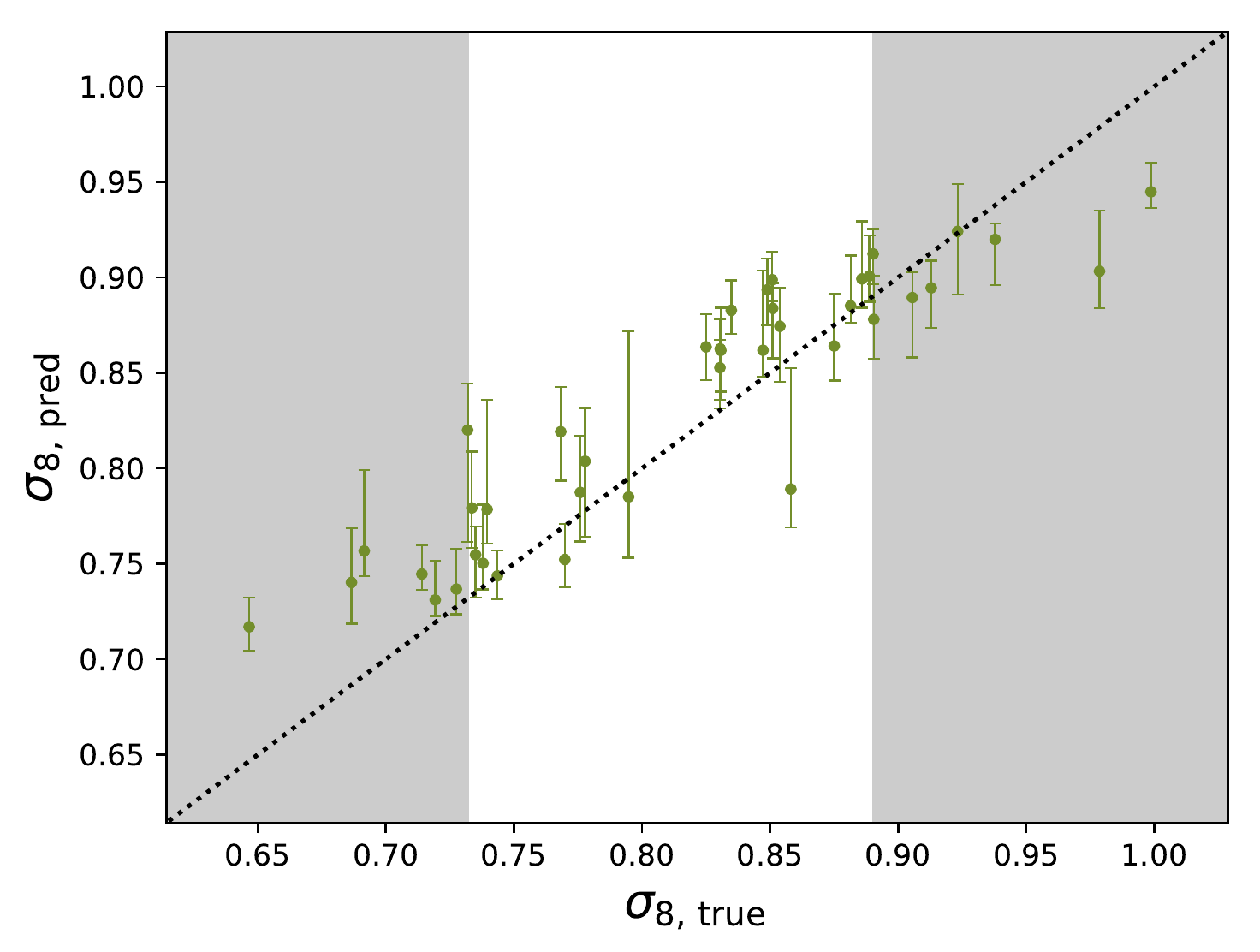} & \includegraphics[width=0.45\textwidth]{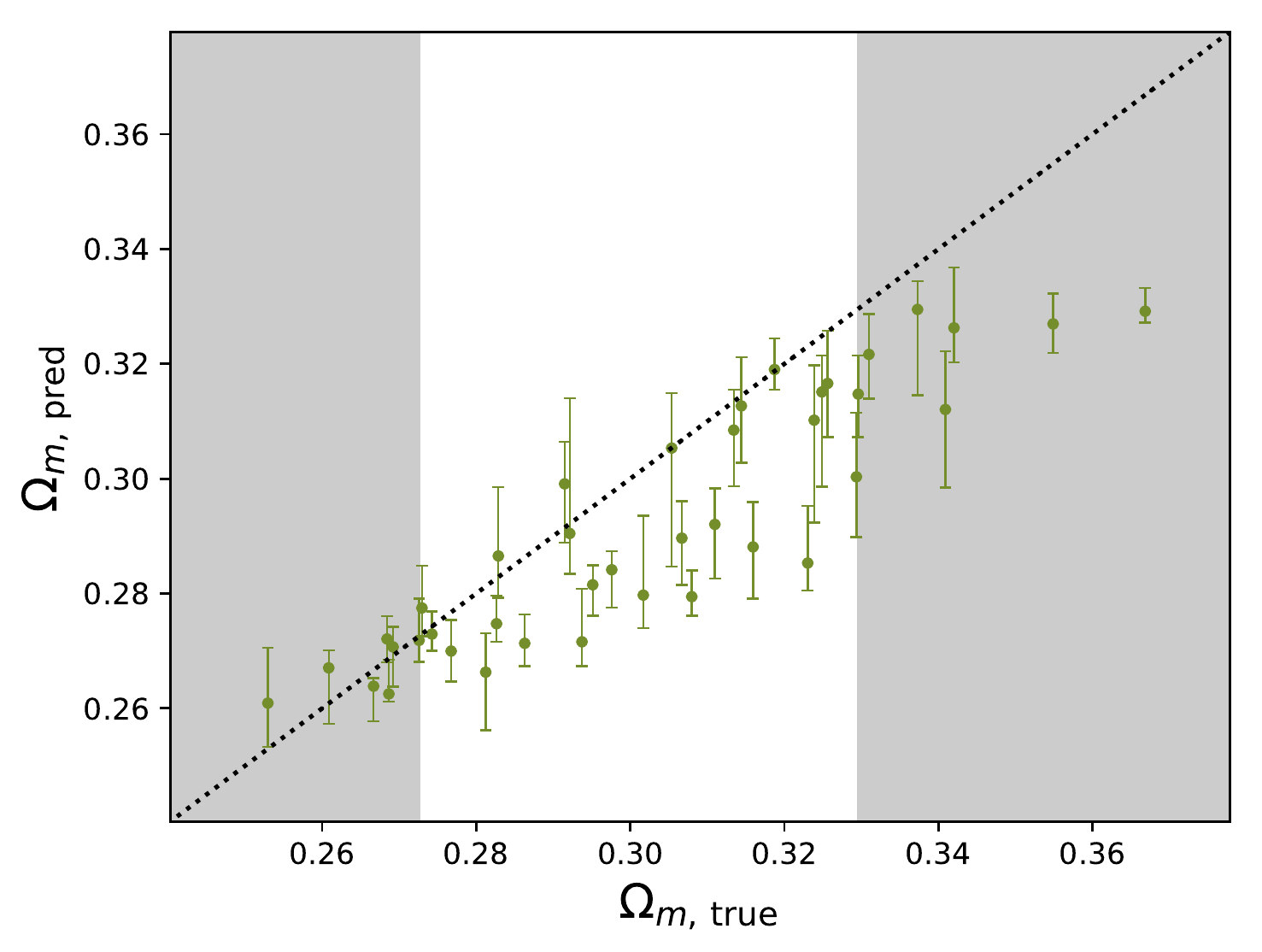} \\
		\includegraphics[width=0.45\textwidth]{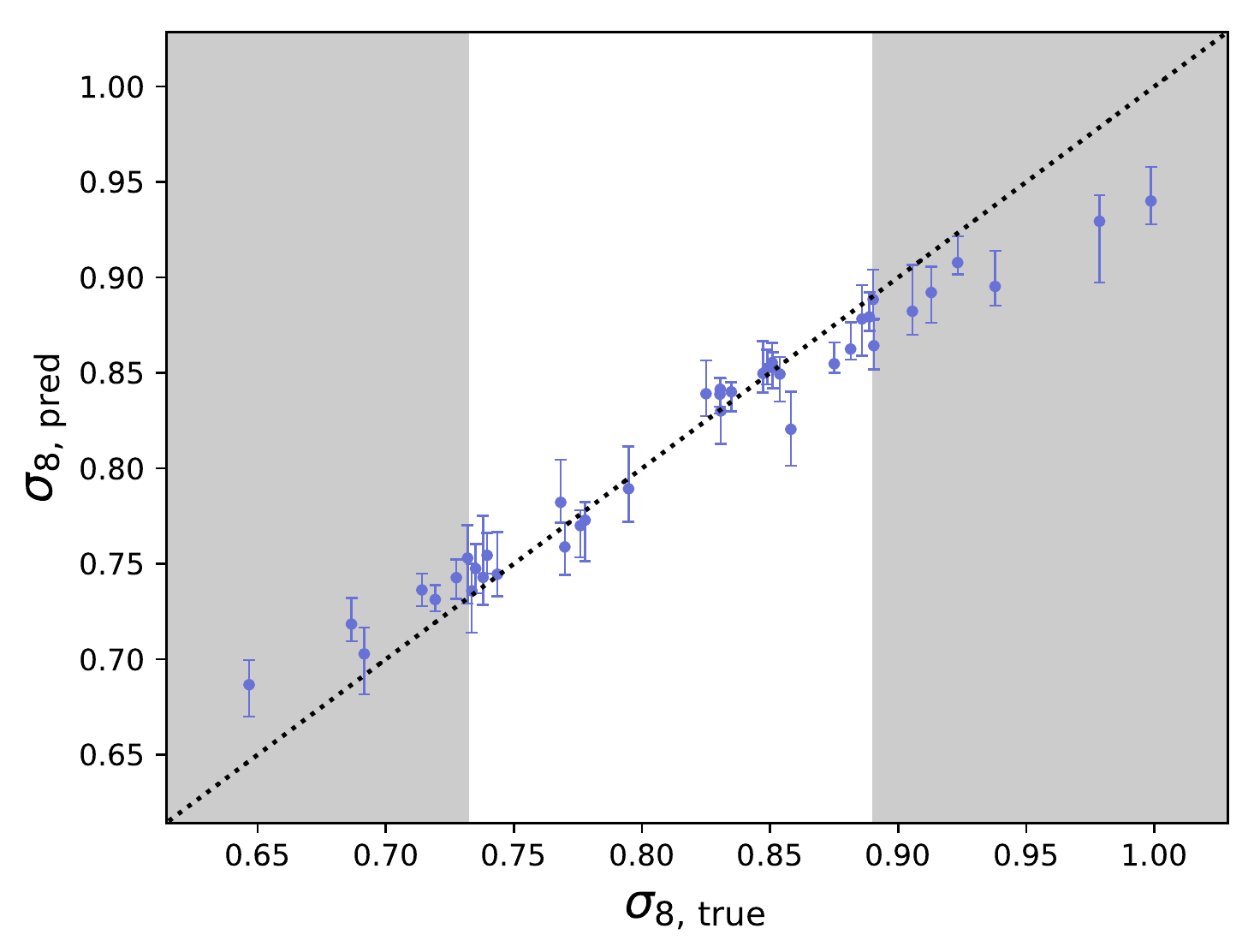} & \includegraphics[width=0.45\textwidth]{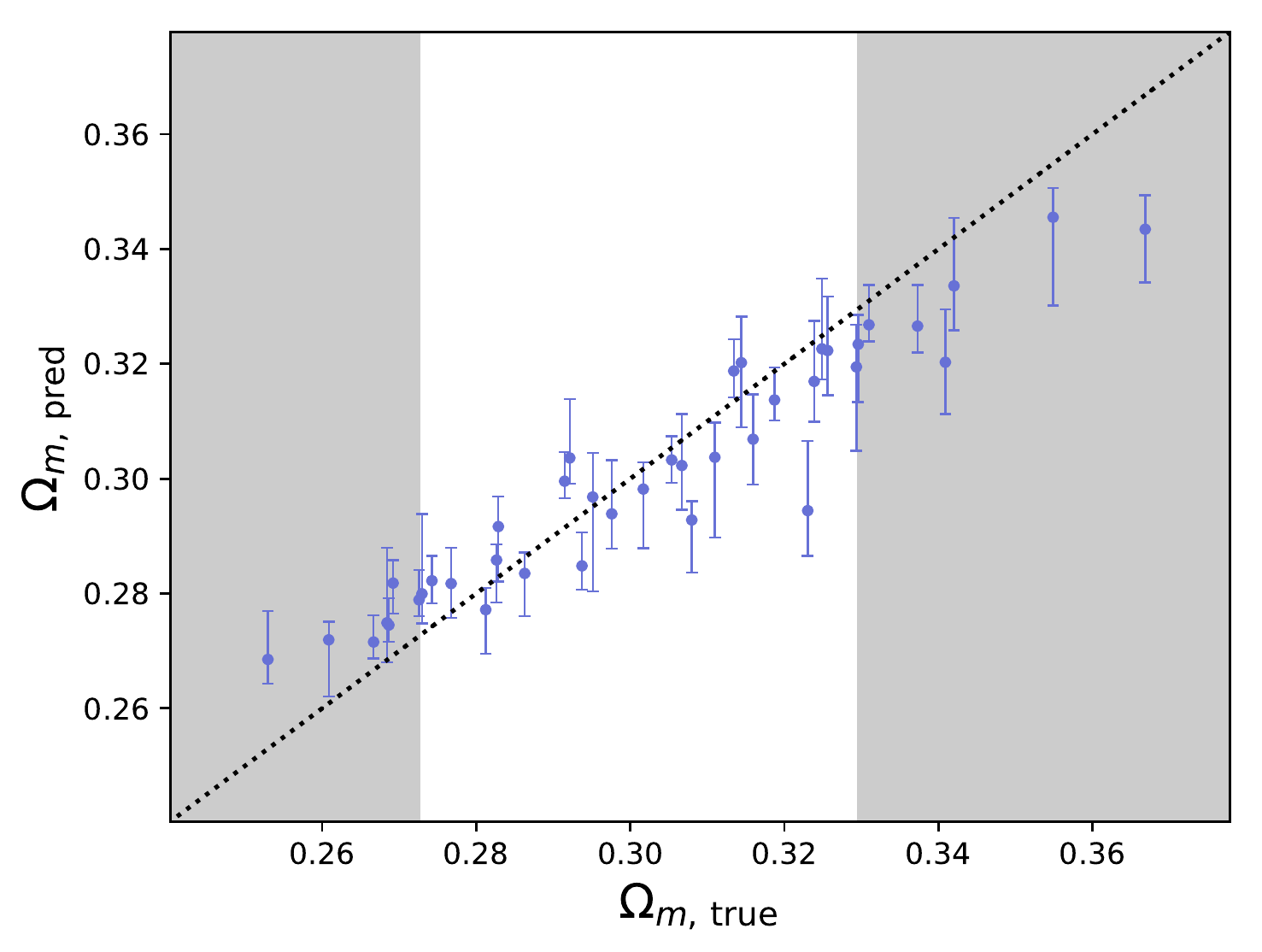} \\
		\includegraphics[width=0.45\textwidth]{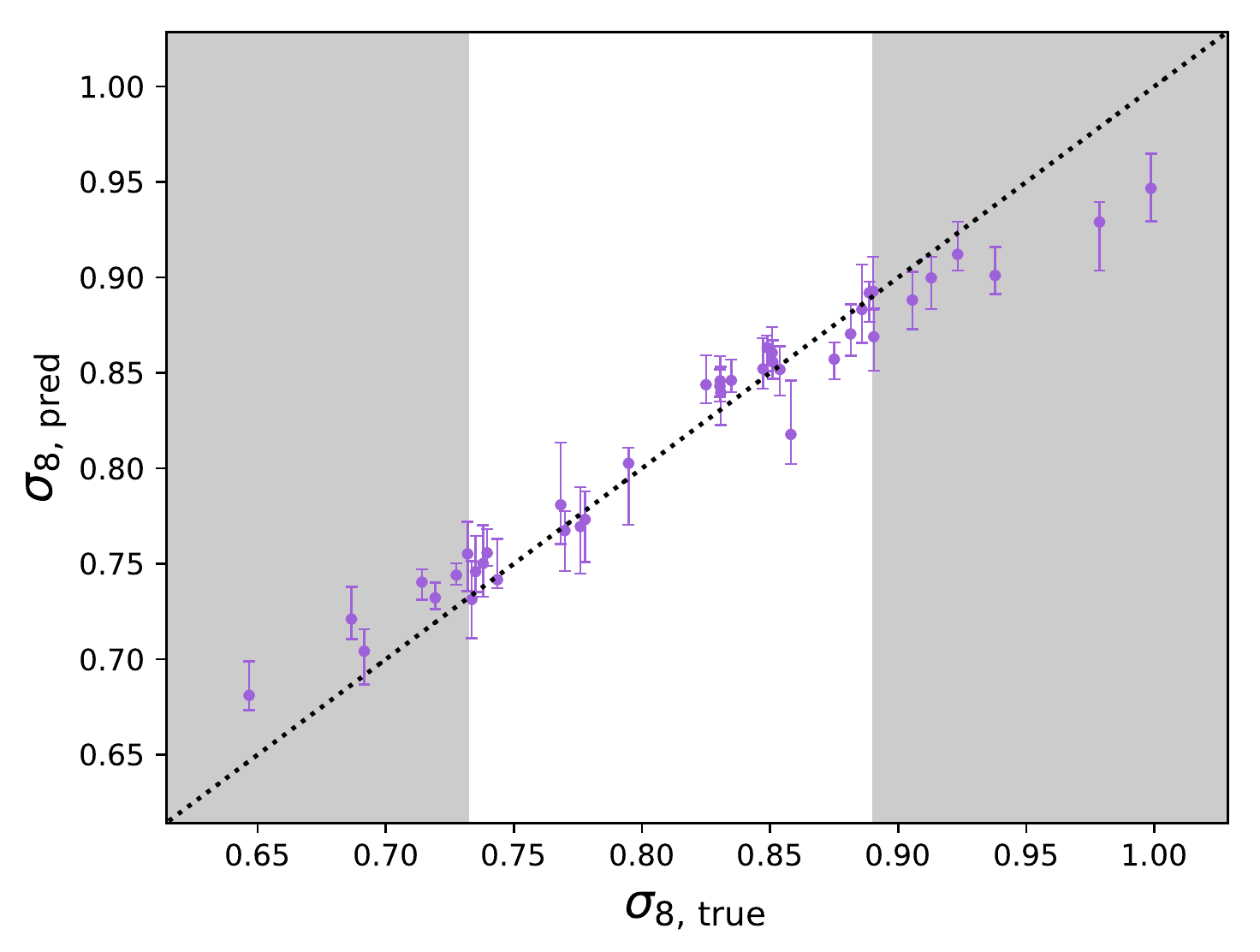} & \includegraphics[width=0.45\textwidth]{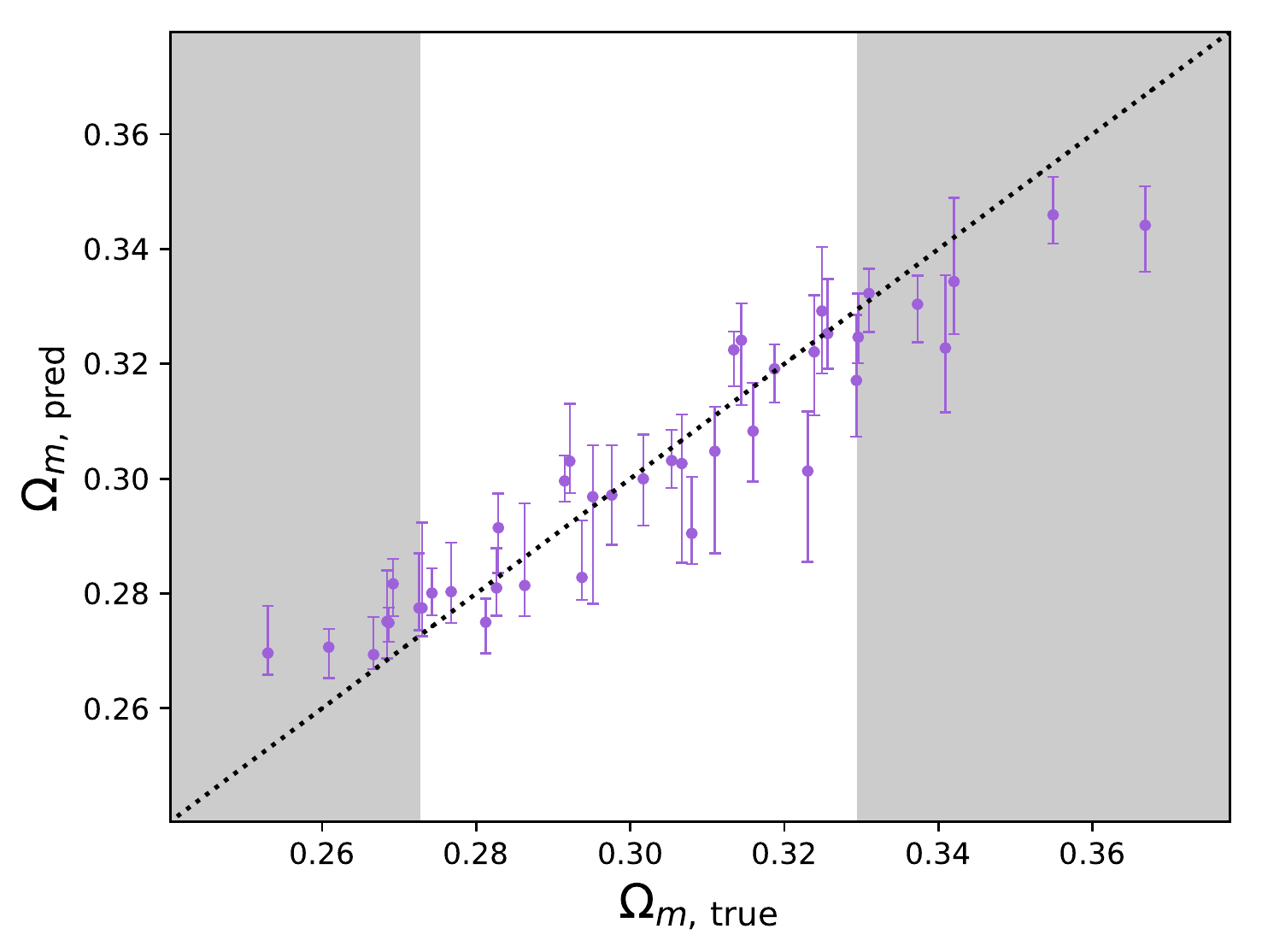} \\
		\end{tabular}
	\caption[]{True and predicted values of \sig (left) and \om (right) for the neural network (NN, green, top), convolutional neural network (CNN, blue, middle) and hybrid CNN (hCNN, purple, bottom).  For the validation data of each of the 40 cosmological models, the median (circles) and middle 68\% (error bars) are shown.  While the predictions typically lie close to the one-to-one line (black dashed) near the central values of \sig and \om, the bias toward the mean is more pronounced at extreme values.  For illustrative purposes, \sig and \om values below the 16th percentile and above the 84th precentile are set against a gray background, while the middle 1-$\sigma$ are shown against a white background.  The CNN and hCNN predictions for the validation set display a significantly tighter scatter than the NN.  This is unsurprising because the NN learns only from the power spectrum (see Figure \ref{fig:ps}), while the CNN and hCNN have more flexibility to learn from the un-preprocessed mock galaxy catalog.}
       	\label{fig:scatter}	
\end{figure*}

\subsubsection{Unbiased Model Selection}
\label{sec:stop}

As highlighted in Figure \ref{fig:biasandslope}, the models do not perform well at extreme values of \sig and \om.  This is unsurprising; machine learning models tend to interpolate much better than they extrapolate.  In practice, one would want to train on a large range of simulated cosmologies extending well beyond a region containing the expected results.  Furthermore, one would expect a bias toward the mean for any cosmology near the edges of the training sample.  Because of this (and for the purposes of model selection only), we limit our analysis to the simulations enclosed in a 68\% ellipse in the \sig-\om plane.\footnote{The selection of simulations used here are shown in a lighter shade of gray in Figure \ref{fig:planck}; the simulations shown in dark gray are near the edges of the \sig-\om plane, are expected to have results that bias to the mean, and are excluded from this particular analysis for this reason.}

In addition to limiting this analysis to the 27 simulations with \sig and \om values closest to the mean cosmology, we also only assess the last 50 epochs of the CNN and hCNN trainings ($0.91<\mathcal{E}\leq 1.0$).  Importantly, we only use the validation data to assess models.  Recall that the training data should not be used in such a way because the model has already explicitly seen this data.  Likewise,  the testing data should not be used to assess models because doing so would unfairly bias the results.

For \textit{each} of the 27 simulations and at each epoch, we calculate the distance between the predicted and true cosmology according to the following:  for each of the 16 validation mock observations per simulation, we predict \sig and \om.  The 68\% error ellipse in the \sig-\om plane is calculated, as is the distance between the true cosmological parameters ($\Omega_{m,\,\mathrm{true}}$ and $\sigma_{8,\,\mathrm{true}}$) and the middle of the ellipse of predicted cosmological parameters ($\Omega_{m,\,\mathrm{mid}}$ and $\sigma_{8,\,\mathrm{mid}}$).  This distance, $\mathcal{Z}$, is calculated according to
\begin{equation}
	\begin{split}
		\mathcal{Z} \, =\,  & \frac{(\Omega_{m,\,\mathrm{true}}-\Omega_{m,\,\mathrm{mid}})\cos{\alpha}+(\sigma_{8,\,\mathrm{true}}-\sigma_{8,\,\mathrm{mid}})\sin{\alpha}}{a^2} +\\
		\\
	                             	   & \frac{(\Omega_{m,\,\mathrm{true}}-\Omega_{m,\,\mathrm{mid}})\sin{\alpha}-(\sigma_{8,\,\mathrm{true}}-\sigma_{8,\,\mathrm{mid}})\sin{\alpha}}{b^2}
	\end{split}
	\label{eq:Z}
\end{equation}
where $\alpha$ is the angle of the best fit 68\% ellipse, $a$ is the length of the semimajor axis, and $b$ is the length of the semiminor axis.  $\mathcal{Z}$, then, is a 2-dimensional z-score, where $\mathcal{Z}=1$ can be interpreted as the true value being on the edge of the 68\% ellipse and $\mathcal{Z}=0$ means that the true and mean predicted values are identical.  We note that this choice favors accuracy over precision because larger error ellipses are more forgiving of large offsets between the predicted and middle predicted cosmological models.

For each epoch, the mean squared error, MSE, as a function of epoch is calculated according to
\begin{equation}
\mathrm{MSE(e)} = \frac{1}{N_\mathrm{sims}}\sum_{i=1}^{N_\mathrm{sims}} \mathcal{Z}^2_i(e)
\label{eq:MSE}
\end{equation}
We select the epoch with the smallest MSE as the final model \textemdash{} and the model least likely to produce biased results \textemdash{} for the CNN and hCNN.  Coincidentally, these ``unbiased'' models are from training epochs that are rather close to each other, epochs 520 and 524 ($\mathcal{E}\approx0.95$) for the CNN and hCNN, respectively.  Selecting, instead, to define a 2-D error ellipse that is averaged over all models and epochs selects the same hCNN model, but prefers a CNN model with marginally tighter error bars and a more significant offset.

Figure \ref{fig:scatter} shows the median and middle 68\% predictions for each of the 40 cosmologies represented in the validation set at these unbiased epochs.  As expected, the model visibly pulls toward the mean for outlying values of \sig and \om.  The CNN and hCNN produce tighter correlations between the true and predicted values than does the NN.

\subsubsection{Planck Testing Set Results}
\label{sec:planck}

\begin{figure}[]
	\begin{centering}
		\includegraphics[width=0.5\textwidth]{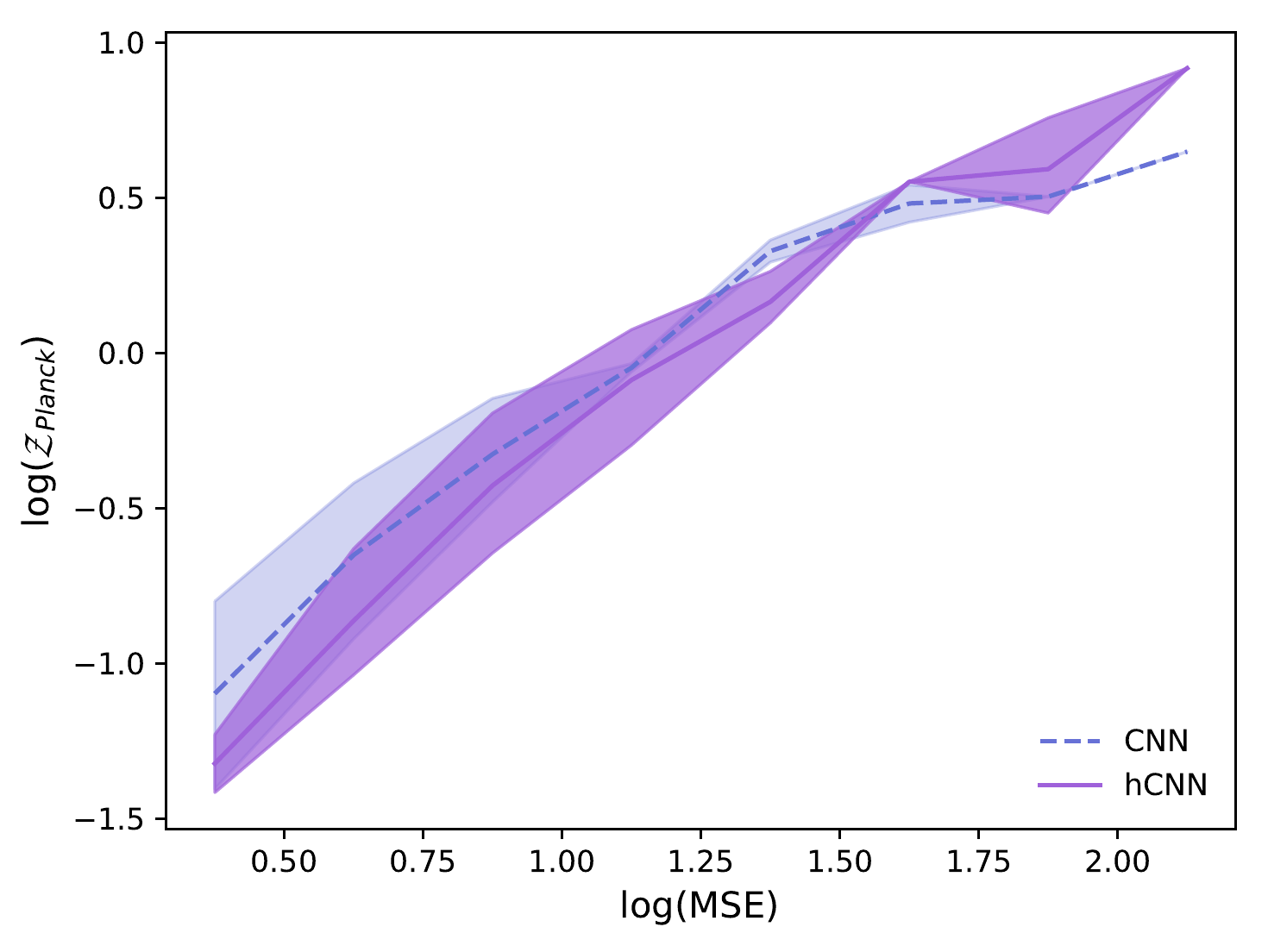}\\
	\end{centering}
	\caption[]{The MSE of the validation set, calculated in equations \ref{eq:Z} and \ref{eq:MSE}, is tightly correlated with the testing set error $\mathcal{Z}_{Planck}$.  Shown are the binned median and 68\% scatter  for the CNN (blue dash) and hCNN (purple solid).  The values tabulated here are restricted to epochs 501-550 ($0.91<\mathcal{E}\leq1.0$).  The tighter correlation between low-MSE and low-$\mathcal{Z}_{Planck}$ models is mildly more pronounced for the hCNN, suggesting that the hCNN might be a more robust approach for producing unbiased results.}
       	\label{fig:MSE}	
\end{figure}

\begin{figure*}[]
	\begin{centering}
		\includegraphics[width=0.8\textwidth]{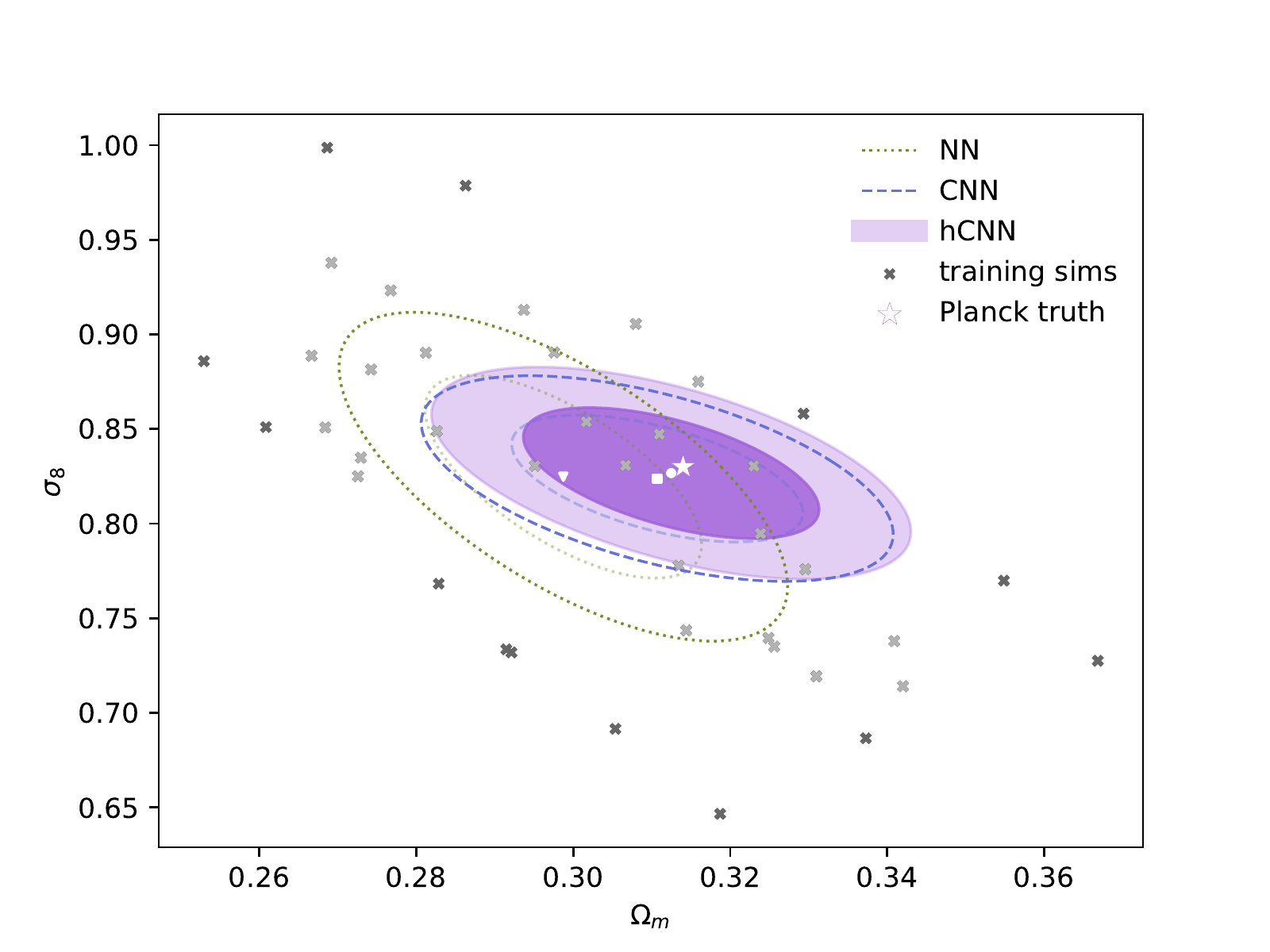}\\
	\end{centering}
	\caption[]{Testing set predictions of the NN (green dotted), CNN (blue dashed), and hCNN (purple filled); shown are the 68\% and 95\% error ellipses.  The NN is heavily influenced by the degeneracy of the training simulations (gray x's) in the \sig-\om plane, and predicts cosmological parameters that are significantly biased toward the mean.  The CNN and hCNN have tighter error ellipses and smaller biases.  The bias toward the mean is mildly smaller for the hCNN (white circle denoting the center of the error ellipse) compared to the CNN (white square).}
       	\label{fig:planck}	
\end{figure*}

Recall that the training set comprises mock observations built from 40 matched-phase cosmological simulations, while the validation set comprises mock observations from a unique portion ($z<200\Mpch$) of those same simulations.  In contrast, the testing set comprises mock observations from non-matched-phase simulations at the \planck cosmology which were populated with galaxies according to an HOD not yet seen by the trained model.  With previously unseen cosmological parameters, HOD, and initial conditions, the \planck testing set is a more fair test of expected error and biases under a realistic set of conditions.  

In the previous section, we posited that the MSE of the validation set might serve as a fair proxy assessment for selecting an unbiased model to apply to an unseen cosmology.  Indeed, the validation MSE and the $\mathcal{Z}$ value for the \planck testing data (denoted $\mathcal{Z}_{Planck}$), are highly correlated, as shown in Figure \ref{fig:MSE}.  The $\log(MSE)$-$\log(\mathcal{Z}_{Planck})$ distribution has a Pearson R correlation coefficient of 0.88 for the CNN and a slightly tighter correlation of 0.93 for the hCNN.  There is no strong evidence of evolution in the MSE-$\mathcal{Z}_{Planck}$ plane as a function of epoch; while low MSE is correlated with low $\mathcal{Z}_{Planck}$, the model is not taking a slow and steady march toward high or low MSE as it trains during epochs 501-550.  The model's loss function should drive a decrease in mean absolute error across the 40 training cosmologies as it trains, while the test shown assesses a different measure of the goodness of fit.

Figure \ref{fig:planck} shows the cosmological constraints for the NN, CNN, and hCNN.  Despite the goodness of training suggested by the results in Figures \ref{fig:err}, \ref{fig:biasandslope},  and \ref{fig:scatter}, the NN never moves beyond predictions that are heavily influenced by the degeneracy of the training simulations.  This is, perhaps, unsurprising.  The power spectrum on which it is trained is calculated from a relatively small volume, $\sim 0.07\,h^{-3}\,\mathrm{Gpc}^3$, in contrast with the effective volume of $\sim6\,\mathrm{Gpc}^3$ of the SDSS DR11 Baryon Oscillation Spectroscopic Survey (BOSS) observation \citep{2015MNRAS.451..539G}.  The volume of the mock observations used in this work is too small to isolate the baryon acoustic peak and reliably measure the acoustic scale. As a result, while the NN predicts \sig in an unbiased way, its predictions for \om are biased very low and pull toward the mean \om of training simulations.

Compared to the NN, the CNN and hCNN predictions are substantially unbiased.  The cosmological constraints in Figure \ref{fig:planck}, as well as the sample of low-$\mathcal{Z}_{Planck}$ models in Figure \ref{fig:MSE} suggest that the vector features included in the hCNN \textit{may} make the model more robust to biasing, though the evidence for the effects of bias as a function of vector features is not strong.

Table \ref{table:summary} tabulates the simulation parameters and testing set results.  For reference, we include the \planck testing set true values; recall that all simulations in the \planck suite of simulations were run at identical cosmologies, so the scatter of these values is $0$.  Table \ref{table:summary} also gives parameters that describe the distribution of the training data for reference.  These include the training set mean \sig, mean \om, and the standard deviation of these, and are used as a benchmark for how the distribution of simulated cosmologies compares to the error bars presented.  

For the trio of ML models, the mean ($\bar{x}$), offset ($\bar{x}-x_\planck$), standard deviation of the predictions (denoted $\sigma$), and 1D $z$-score (offset$/\sigma$) are also given.  The NN is the most biased of the trio, particularly in \om, with the mean prediction $\sim$1.3-$\sigma$ away from the true value.  From the bias and error bars associated with the NN, we can conclude that the box volume is likely not large enough for the power spectrum to be diagnostic.  Moving to a larger mock observations that can more reliably measure the acoustic scale is likely to improve the NN technique.

The CNN and hCNN, on the other hand, both predict \sig to within 3\% and \om to within 4\%.  The CNN and hCNN error bars are similarly sized, but the hCNN exhibits a bias that is smaller than the CNN by about a factor of $2$.  However, the bias in both the CNN and hCNN are small, and further studies on larger mock observations are needed to make strong claims about the potential de-biasing advantage of the hCNN architecture.

\begin{deluxetable*}{l r r r r r r r r}

\tablecaption{{Results Summary}\label{table:summary}}
\tablehead{ & \multicolumn{4}{c}{\sig}& \multicolumn{4}{c}{\om}\\
 \colhead{} 	
 & \colhead{mean} & \colhead{offset} &  \colhead{$\sigma$} & \colhead{$z$} 
 & \colhead{mean}& \colhead{offset} & \colhead{$\sigma$} & \colhead{$z$}}

\startdata
Training Set & $ 0.818 $  & \nodata & $ 0.083 $  & \nodata  & $ 0.303 $  & \nodata & $ 0.027 $  & \nodata \\
\planck Testing Set & $ 0.830 $  & \nodata & \nodata  & \nodata   & $0.314$  & \nodata  & \nodata  & \nodata \\
\tableline
NN & $ 0.825 $  & $ 0.005 $  & $ 0.035 $  & $ 0.147 $   & $ 0.299 $  & $ 0.015 $  & $ 0.012 $  & $ 1.307 $  \\
CNN & $ 0.824 $  & $ 0.006 $  & $ 0.022 $  & $ 0.278 $   & $ 0.311 $  & $ 0.003 $  & $ 0.012 $  & $ 0.268 $  \\
hCNN & $ 0.827 $  & $ 0.003 $  & $ 0.023 $  & $ 0.144 $   & $ 0.312 $  & $ 0.002 $  & $ 0.012 $  & $ 0.121 $  \\
\enddata
\label{table:summary}
\end{deluxetable*}

\section{Discussion \& Conclusion}
\label{sec:conclusion}

We have presented a trio of ML approaches for learning \sig and \om from a mock 3D galaxy survey. 
The neural network (NN) uses the binned power spectrum as input, and is processed through a fully connected neural network architecture.  The convolutional neural network (CNN) uses a spatially binned 3D galaxy distribution; this is processed through a series of convolutions and pooling, and finally through a fully connected network.  The hybrid CNN (hCNN) merges the two.

The methods are trained and tested on a sample of  mock surveys are built on the \abacus suite of cosmological $N$-body simulations, and the mock surveys include a variety of galaxy formation scenarios through the implementation of generalized halo occupation distributions (HODs).  The full training sample spans a large parameter space \textemdash{} 6 cosmological parameters and 6 HOD parameters.

We describe a number of best practices for preventing the a 3D CNN or 3D hCNN from memorizing structure and producing overly-optimistic results on the validation data.  Most important is setting aside an independent portion of \textit{all} simulations as a validation set to assess the goodness of fit.  This validation set should ideally drawn from the same portion of the box to prevent the deep network from memorizing correlated structure across simulations stemming from simulations with matched initial phases. Other best practices include recentering the box, aggressive pooling to restrict the models' knowledge of slab-size length scales, subsampling the galaxy catalog to prevent the model from learning from the aggregate number of galaxies within a volume, and employing the standard suite of axial flips and rotations to account for rotational invariance.

We have shown that the validation set MSE is a useful proxy for selecting a model that will produce unbiased estimates of the cosmological parameters, even when presented with previously unseen cosmological and HOD parameters.

The model is limited by the availability of simulated data:  it is trained and tested on relatively small volumes ($\sim 0.07\,h^{-3}\,\mathrm{Gpc}^3$, which is 1/20 of the simulation box volume).  Furthermore, we train with only 40 training simulations at a variety of cosmologies that vary in $\Omega_{CDM}\,h^2$, $\Omega_b\,h^2$, \sig, \Ho, $w_0$, and $n_s$, which have been populated with galaxies according to a flexible HOD with 6 parameters.  Yet, even within these limitations \textemdash{} the small volumes and large cosmological and HOD parameter space \textemdash{} we have shown that it is possible to robustly train a model that can learn 
\sig and \om directly from a catalog of galaxies. 

Developing more realistic mock observations that span the cosmological and galaxy formation parameter space is an essential next step for applying 3D hybrid CNNs to observational data.  These extensions to the existing mock observations include adopting more diversity in cosmological parameters, taking advantage of larger training mock observations, employing additional flexibility in galaxy models, and modeling real survey embeddings.  As such training data become available, 3D hybrid CNNs have the potential to become a powerful tool for extracting cosmological information from next-generation spectroscopic surveys.\\ \\

\acknowledgments{We thank Alexei Efros, Melanie Fernandez, Zolt\'an Haiman, Paul La Plante, Jos\'e Manuel, Szymon Nakoneczny, Junier Oliva, Barnab\'as P\'oczos, Siamak Ravanbakhsh, Dezs{\"o} Ribli, Alexey Vikhlinin, and Javier Zazo for their helpful feedback on this project.}\\ \\

\bibliography{references}
\bibliographystyle{apj}

\appendix

\section{On the Life Cycle of CNNs}
\label{sec:lifecycle}

CNNs are traditionally trained to minimize a loss function such as mean squared or absolute error, yet it is not obvious that this is an ideal approach for astronomical and cosmological applications.  In this section, we present more on the life cycle of our CNN and show additional plots that have been useful in interpreting fits and designing our two-phase training scheme.

While figures showing traditional metrics can be diagnostic, they can be difficult to interpret for models that regress more than one parameter.  Such traditional figures include error as a function of epoch (e.g., Figure \ref{fig:err}) and 1-to-1 scatter of true and predicted values (e.g., Figure \ref{fig:scatter}).  It is concerning that typical early stopping routines rely on these test statistics to determine when a model is well fit because using such diagnostics blindly can lead to unexpected or overpessimistic results.

\begin{figure*}[b]
	\begin{centering}
	\begin{tabular}{c c}
		\includegraphics[width=0.5\textwidth]{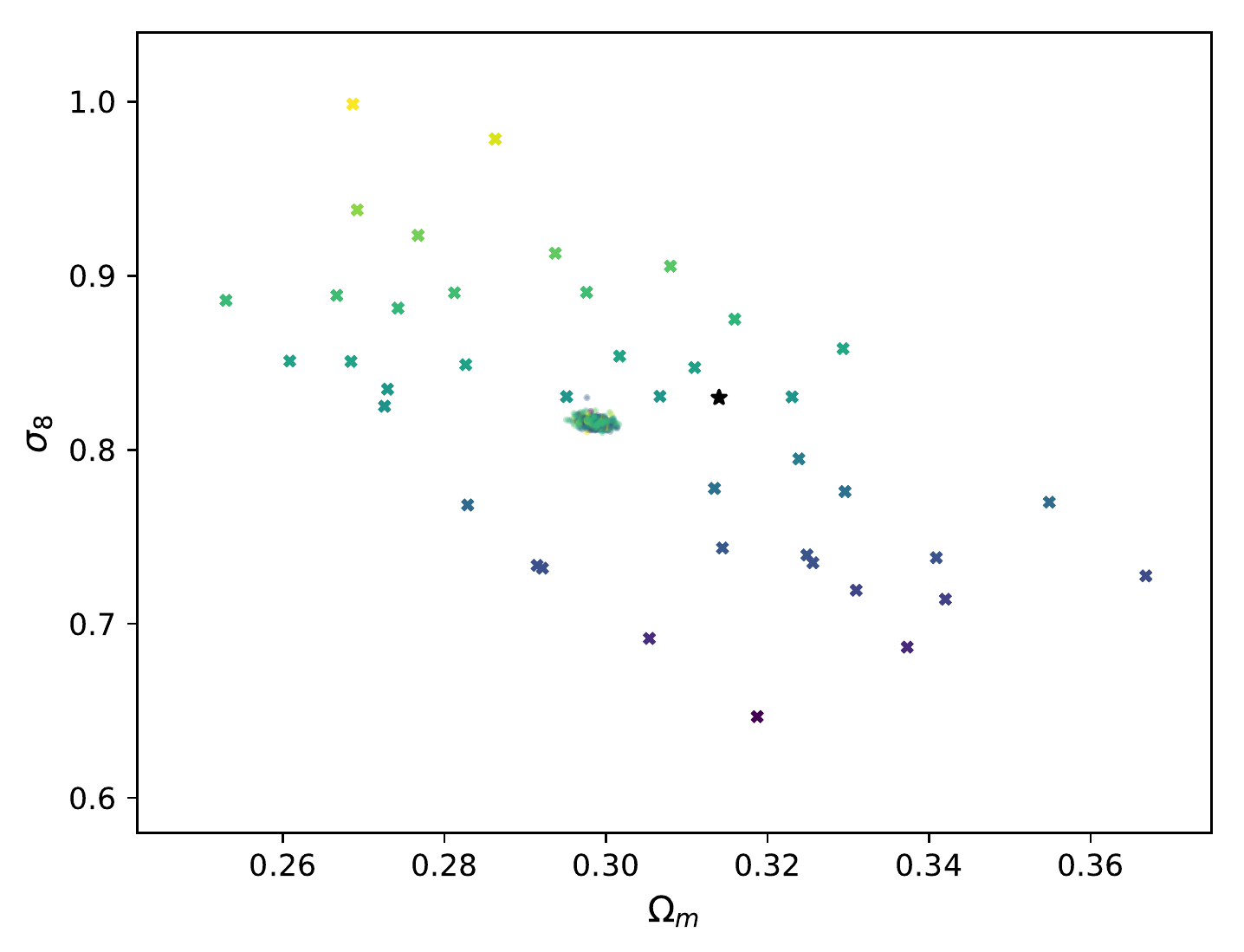}&\includegraphics[width=0.5\textwidth]{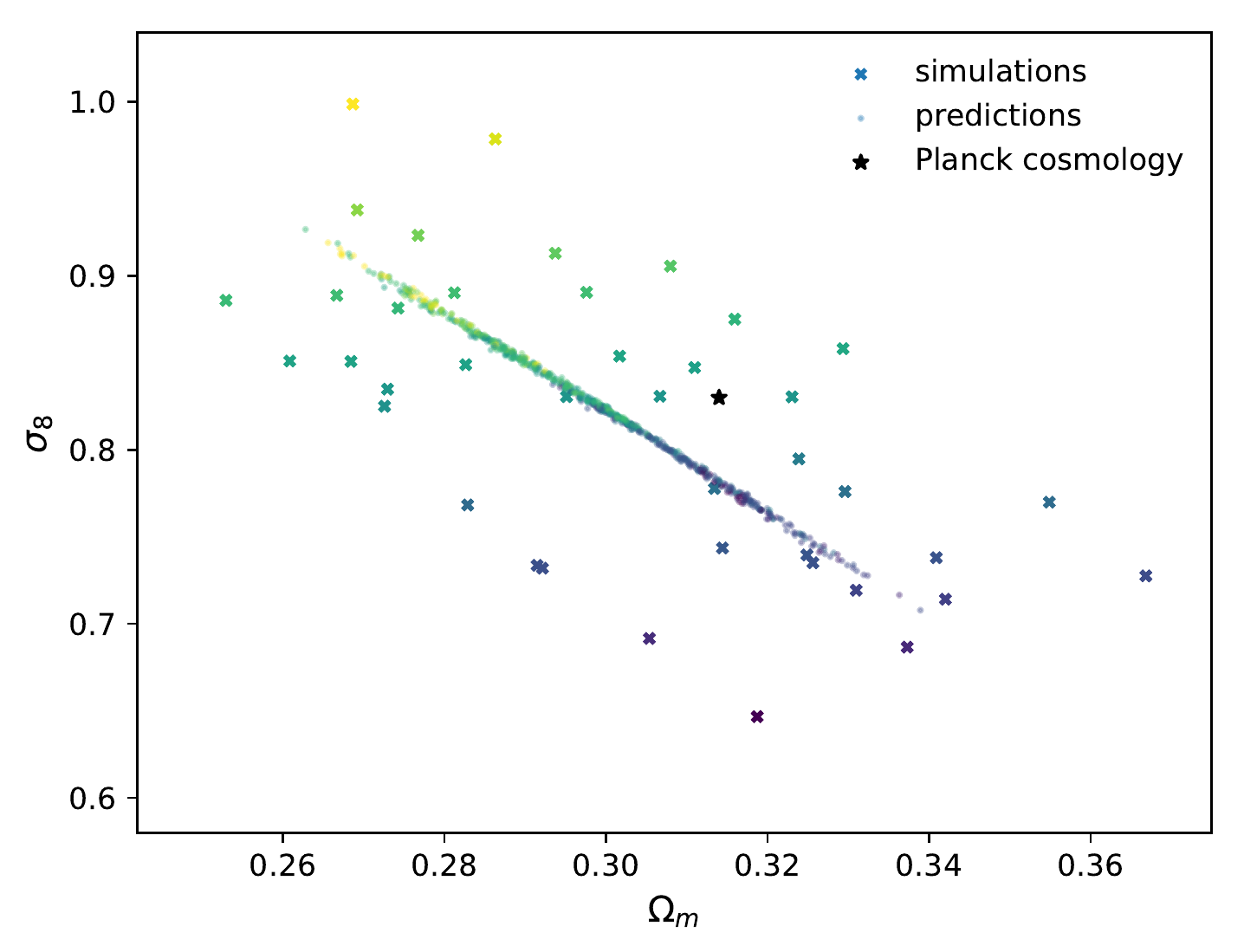}\\
		\includegraphics[width=0.5\textwidth]{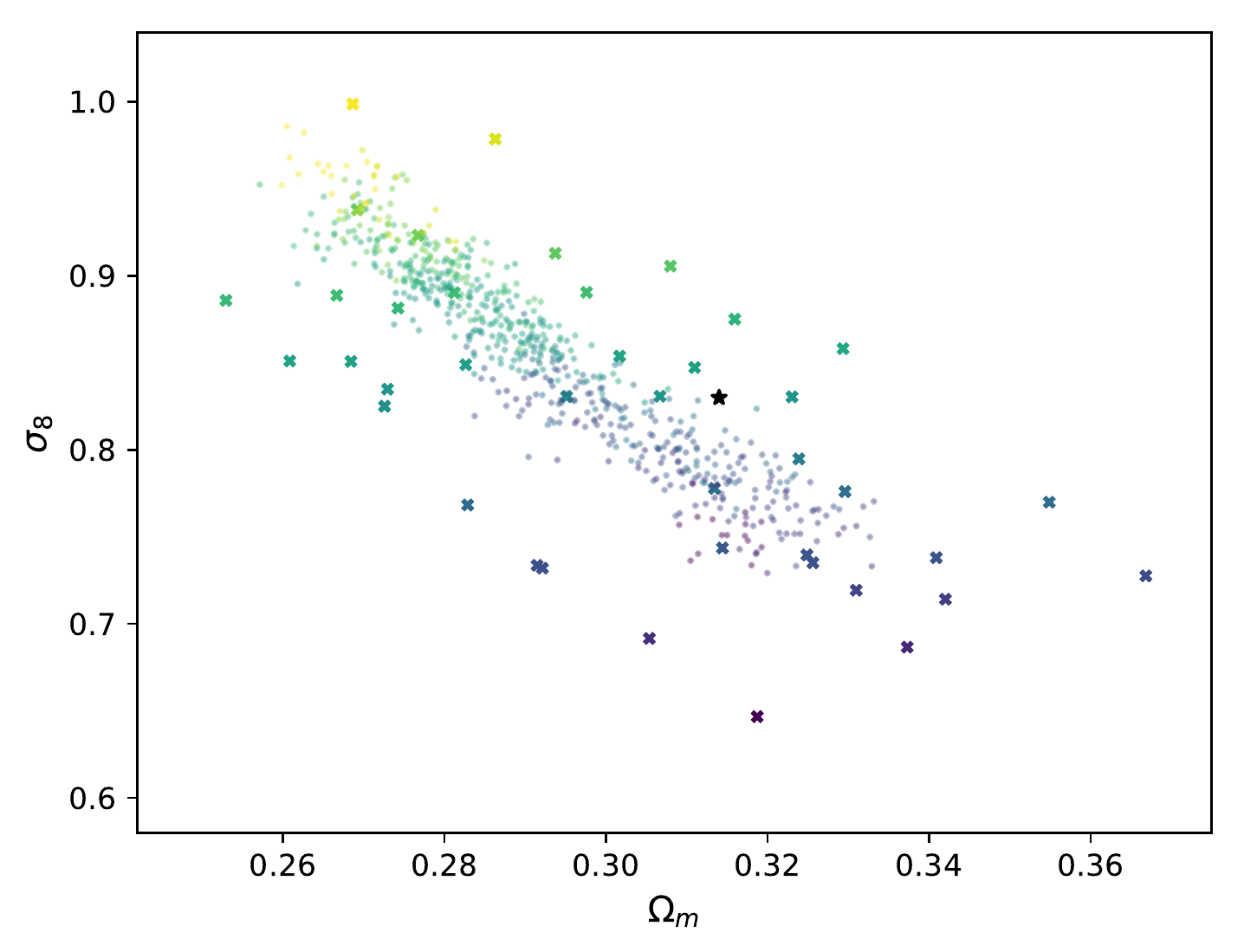}&\includegraphics[width=0.5\textwidth]{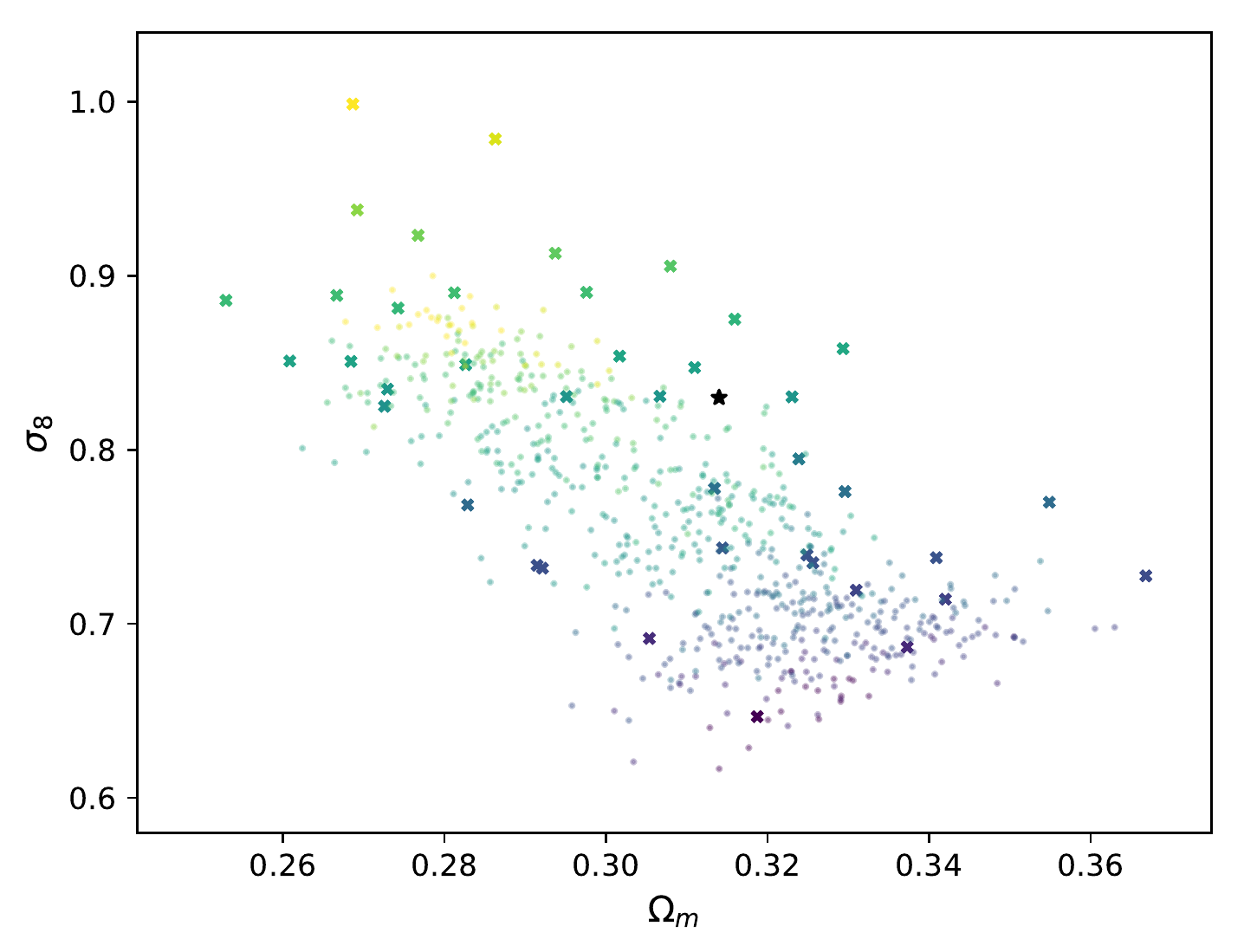}\\
		\includegraphics[width=0.5\textwidth]{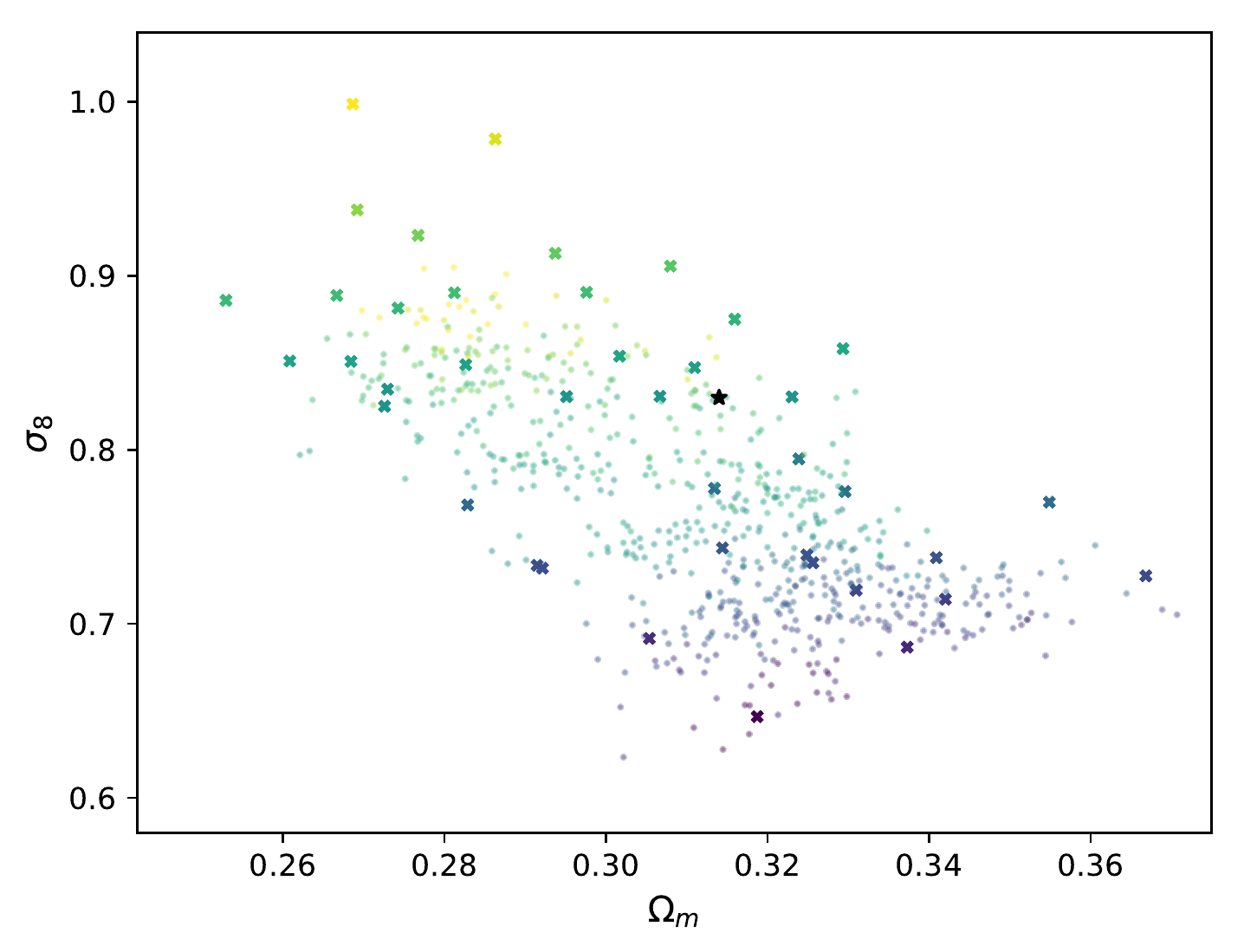}&\includegraphics[width=0.5\textwidth]{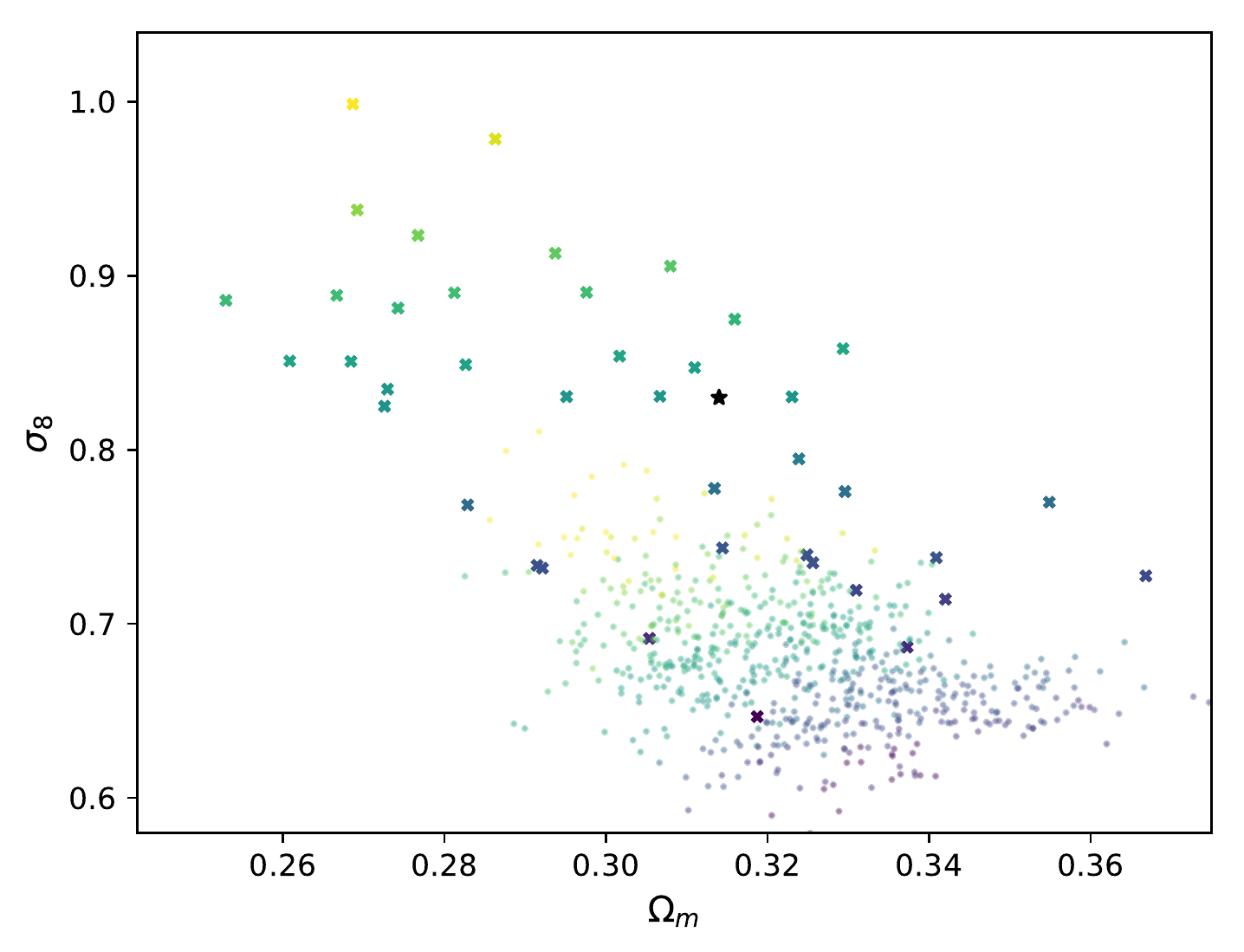}\\
	\end{tabular}
	\end{centering}
	\caption[]{The life cycle of CNNs.  Training data (crosses) are colored according to their \sig values, and predictions on the validation data (circles) are likewise colored according to their \emph{true} (not predicted) \sig values.  Early in training, the model learns reasonable values for \sig and \om, eventually learning a tight degeneracy in this space, and finally achieving a more diverse representation of the simulations.  Shown are, from top left to bottom right, epochs 5, 30, 100, 175, 219, and 220 in phase 1 of the training scheme. }
       	\label{fig:cnn_thisisyourlife}	
\end{figure*}

Figure \ref{fig:cnn_thisisyourlife} shows the validation data 2D predictions as a function of epoch.  Unsurprisingly, at epochs as early as 5, the model has learned to predict a mean value but cannot differentiate among models.  This is encouraging and expected; the model, which is initialized to completely random weights and biases, learned a reasonable values for \sig and \om in the first few epochs.

The model predictions at epoch 30, though, are a surprising extension of this predition of the mean.  In Figure \ref{fig:err}, the error as a function of epoch slowly and steadily decreases for the first few epochs, then begins to oscillate.  At epoch $\sim30$, this initial plunge has come to an end, and an error-based early stopping scheme might suggest that these results are sufficient.  A one-to-one plot of true and predicted \sig and \om will tell a similar story \textemdash{} the results bias toward the mean and the scatter is larger than is to be desired, but the model has clearly learned trends in the data and a diversity of \sig and \om values.  Yet, when viewed as a scatter plot in the \sig-\om plane (in the top right corner of Figure \ref{fig:cnn_thisisyourlife}), it is clear that the CNN has learned a 2D version of predicting the mean:  it has produced predictions that spread along the degeneracy direction of the training simulations, with the predictions arranged in a sensible way (i.e.~the predictions of the high-\sig simulations are indeed at high \sig).  

It is only by delving into a ``high error'' regime that the CNN starts to make progress beyond this tight degeneracy.  Between epochs 30 and 175, we see large oscillations in MSE.  Epoch 100 is shown as an example epochs in this region.  Despite the fact that Figure \ref{fig:err} shows error increasing and oscillating in this epoch range, something important and meaningful is happening under the surface.  The model is starting to produce more diversity in predictions, expanding the range of predictions in the direction orthogonal to the degeneracy of the simulations.  At epoch 175, the predictions are still biased toward the mean, but at least span a wider spectrum of possibilities.

Here, we can take an alternate timeline and continue with phase 1 of the training scheme for a few more epochs.  Recall that, in the training scheme presented in the main text, we transition to a lower learning rate at epoch 175.  At epochs 219 and 220 in this alternate timeline, we begin to see the oscillations in bias.  While the results for epoch 219 look reasonable, the results for epoch 220 are biased to very low \sig; such large swings in biases hint that the step size is too large.

Another alternative timeline transitions from phase 1 (high learning rate) to phase 2 (lower learning rate) as early as epoch 30, with disastrous results.  The epoch 30 model has not yet learned much beyond the degeneracy of the simulations, and when it is moved to a much smaller learning rate, it fails to learn a diversity of predictions in the \sig-\om plane, instead producing predictions along a tight curve for many epochs.

While they are certainly valuable, traditional methods for understanding how well a CNN has fit can be difficult to interpret, particularly when assessing models trained to predict multiple parameters.  Employing early stopping routines that assess a single statistical measurement of error can lead to models that have not yet learned a range of predictions in the parameter space.  Appropriately assessing the diversity of predictions, identifying epochs to stop training, and developing intuition for training deep models will an essential step toward properly using these powerful tools in astronomical and cosmological applications. 

\end{document}